\documentstyle[12pt,emulateapj,apjfonts,psfig]{article}

\tighten

\lefthead{Briggs et al.}
\righthead{BATSE GRB Location Errors}

\begin{document}

\submitted{1999 January 8}

\title{The Error Distribution of BATSE GRB Locations}

\author{
Michael S. Briggs \altaffilmark{1},
Geoffrey N. Pendleton \altaffilmark{1},
R. Marc Kippen \altaffilmark{2},
J. J. Brainerd \altaffilmark{1}, 
Kevin Hurley \altaffilmark{3},
Valerie Connaughton \altaffilmark{4},
Charles A. Meegan \altaffilmark{5}
}

\altaffiltext{1}{Department of Physics, University of Alabama in Huntsville,
Huntsville, AL 35899}
\altaffiltext{2}{Center for Space Plasma, Aeronomic and Astrophysics Research, 
University of Alabama in Huntsville,
Huntsville, AL 35899}
\altaffiltext{3}
{Space Sciences Laboratory, University of California, Berkeley, CA 94720-7450}
\altaffiltext{4}{NRC \& NASA/Marshall Space Flight Center, Huntsville, AL 35812}
\altaffiltext{5}{NASA/Marshall Space Flight Center, Huntsville, AL 35812}

\begin{center}
To appear in {\it The Astrophysical Journal Supplement Series} \\
\copyright \hspace{0.05mm} 1999 by the American Astronomical Society.
\end{center}

\begin{abstract}

Empirical probability models for BATSE GRB location errors are developed
via a Bayesian analysis of the separations between BATSE GRB locations and
locations obtained with the InterPlanetary Network (IPN).
Models are compared and their parameters estimated using
392 GRBs with single IPN annuli and 19 GRBs with intersecting IPN
annuli.
Most of the analysis is for the 4Br BATSE catalog;  earlier
catalogs are also analyzed.
The simplest model 
that provides a good representation of the error distribution
has 78\% of the probability in a `core' term
with a systematic error of 1.85 degrees and the remainder in an extended
tail with a systematic error of 5.1 degrees, implying a
68\% confidence radius for bursts with negligible statistical uncertainties
of 2.2 degrees.
There is  evidence for a more complicated model in which the
error distribution depends on the BATSE datatype that was used to obtain the
location.
Bright bursts are typically located using the CONT datatype, and according
to the more complicated model, the 68\% confidence radius for CONT-located
bursts with negligible statistical uncertainties is 2.0 degrees.

\end{abstract}


\section{Introduction}

An improved model for the distribution of errors
for gamma-ray burst (GRB) locations obtained with the
Burst and Transient Source Experiment (BATSE) is presented.
The error model should aid in using the BATSE locations in projects
such as searching for counterparts
and searching for evidence of repetition and clustering, including
gravitational lensing.
Most of the analysis  applies to the 4Br catalog 
\markcite{Pac98}
(Paciesas et al. 1998), which has some revised locations compared to
earlier catalogs, including the initial 4B catalog 
\markcite{Pac97} (Paciesas et al. 1997),
which was released electronically and on CDROM.
Summary results are presented for earlier catalogs.

The BATSE instrument consists of eight modules located on the corners of
the Compton Gamma-Ray Observatory (CGRO).    Each module contains a
Large Area Detector (LAD), consisting of a 1.27 cm thick by 50.8 cm diameter
NaI crystal viewed by three photomultiplier tubes.    Each LAD has an
approximately cosine response.    GRB locations are determined with the
program LOCBURST by modeling
the responses of either four or six detectors as a function of assumed
source location, intensity and spectrum 
\markcite{Pen98} (Pendleton et al. 1998).  
The detector response model is based upon Monte Carlo photon propagation,
laboratory measurements and space observations.   It includes scattering
from nearby spacecraft structures and from the earth's atmosphere 
\markcite{Pen95,Pen98} (Pendleton
et al. 1995, 1998).

The error  models presented in this paper are obtained 
empirically by statistical
comparison of BATSE locations with locations obtained by other techniques.
The ideal reference dataset would have many locations to enable a good
statistical comparison and would consist of locations with errors significantly
smaller than the BATSE errors so that all discrepancies could be attributed
to BATSE. 
If such a dataset existed, there would probably be little interest in
the BATSE locations.
The reference set we use is the locations of the InterPlanetary Network
(IPN) Supplements to
the BATSE 4Br catalog 
\markcite{KH98a,KH98b,Lar97,Lar98,KH_98}
(Hurley et al. 1998a,b; Laros et al. 1997,1998; Hurley 1998).               
Although this dataset is large, most of the
IPN locations are highly accurate only in  one dimension.

\section{Analysis Approach}

\subsection{The Problem}

The goal is an improved model of the error distribution of BATSE GRB locations.
The error model is a probability density function $p$ defined on the sphere.
A probability density function can have any nonnegative value;
integrating $p$ with respect to solid angle element $d \Omega$
over any solid angle
region $\Omega$  yields the probability $P$ (between 0 and 1)
that the true location is in the region $\Omega$.

A portion of
the errors originates from the Poisson fluctuations
of the detected counts.   These fluctuations are propagated into
an estimated statistical uncertainty $\sigma_{\rm stat}$ 
by the program LOCBURST.   
The locations are found by LOCBURST by $\chi^2$-minimization, and
$\sigma_{\rm stat}$ is calculated as the radius of the 
spherical small circle with the
same area as the 68\% confidence ellipse obtained from the Hessian
matrix of derivatives of $\chi^2$ 
\markcite{Pen98} (Pendleton et al. 1998).
In this paper we assume that the total error distribution
is azimuthally symmetric.
The value of $\sigma_{\rm stat}$ for each GRB
is listed in the 4Br catalog 
\markcite{Pac98} (Paciesas et al. 1998), or, for bursts
with unchanged locations, in the 3B catalog 
\markcite{Mee96} (Meegan et al. 1996).
We attribute the difference between the
total error $\sigma_{\rm tot}$ and the statistical 
uncertainty $\sigma_{\rm stat}$
to a systematic error $\sigma_{\rm sys}$.
Possible sources of systematic errors include:
inaccuracies in the assignment of energies to channel boundaries,
inaccuracies in the deadtime correction of high count rates,
the approximations of the statistical and total error boxes as circles,
inaccuracies in the detector response model as a function of photon energy
or direction,
including inaccuracies in the model of scattering from the spacecraft and
the earth's atmosphere, 
deviations of the actual GRB spectra from the power law assumed in most cases,
and
inaccuracies in background subtraction.

The first published BATSE catalog, the 1B catalog 
\markcite{Fish94} (Fishman et al. 1994),
had 260 GRBs.   
The error estimate was based on eleven events with small error boxes
from IPN data.
A root-mean-square difference between the actual BATSE-IPN
separations and the BATSE statistical uncertainties estimated a systematic error
of $\sigma_{\rm sys} = 4^\circ$ 
\markcite{Fish94} (Fishman et al. 1994).
The 3B catalog 
\markcite{Mee96}
(Meegan et al. 1996) introduced an improved location algorithm.
The analysis of the 3B location errors expanded the reference dataset to include
38 locations obtained with CGRO/COMPTEL or WATCH in addition to 12
IPN locations.
The systematic  error $\sigma_{\rm sys}$ was estimated to be $1.6^\circ$
from the root-mean-square difference between the actual separations to
these 50 reference locations and the BATSE statistical uncertainties.
The total errors of the COMPTEL or WATCH locations are not much smaller than
the BATSE errors; in this situation a very good understanding of the errors
of the reference locations is required, otherwise the portion of the 
location discrepancies due to BATSE may be miscalculated.
An overestimation of the COMPTEL or WATCH location errors may have caused 
the BATSE systematic error to be somewhat underestimated.
In addition, we show that the BATSE error distribution is not well
characterized by a single-valued systematic error so that various
techniques of estimating a single systematic error yield different values
\markcite{Pen98} (see, e.g., Pendleton et al. 1998).

We term the  simple error model specified in the 3B paper 
\markcite{Mee96} (Meegan et al. 1996),
that the total error $\sigma_{\rm tot}$
is formed by adding
the statistical uncertainty $\sigma_{\rm stat}$ and 
a systematic error $\sigma_{\rm sys}$ of $1.6^\circ$
in quadrature, the ``minimal'' model.

A rough rule of thumb learned in this analysis is that with $N$ reference
locations, models with up to $\sim N/50$ parameters may be fit.
Exceeding this empirical
limit typically results in models difficult or impossible to
fit, and, if a fit is obtained, nonsense parameter values.
There are only about 20 `point' locations available from IPN data,
but by using a more sophisticated analysis method 
\markcite {GL96} (Graziani and Lamb 1996),
the several hundred single annuli in the IPN Supplements
may be used to constrain the BATSE location error distribution.
Formerly representations of the systematic error were limited to using
a single parameter;  considerably more complicated error models can
now be tested.

\subsection{Reference Data}

The IPN location technique compares the arrival times of the GRB wavefront
at widely separated instruments, thereby determining the angles the
plane wave makes with the vectors connecting each pair of instruments
\markcite{KH98a} (Hurley et al. 1998a).
With only one pair of spacecraft, one angle is determined, and the error region
is an annulus.   This is the typical case during the time period of the 
4B catalog.    Ideally, three or more spacecraft have separations of
inter-planetary scale, multiple annuli are found, and their intersections
give a small error box.
The IPN Supplements to the 4Br catalog contains 458 annuli for 412 GRBs.
Of these 412 GRBs, 366 were observed by only one deep-space instrument,
{\it Ulysses} \markcite{KH98a,KH98b} (Hurley et al. 1998a,b).
The remaining 46 GRBs were observed by {\it Ulysses} and another deep-space 
instrument such as Pioneer Venus Orbiter or Mars Observer
\markcite{Lar97,Lar98} (Laros et al. 1997,1998).

Measurements of the actual separations $\gamma$ are available for the few
cases with small error boxes from intersecting IPN annuli.
For the cases for which
only single annuli are
available, the BATSE-IPN separations are characterized by the
distances of closest approach $\rho$
of the annuli to the BATSE locations (Fig.~1).

\begin{figure*}[tbp!]
\begin{center}
\mbox{
\psfig{figure=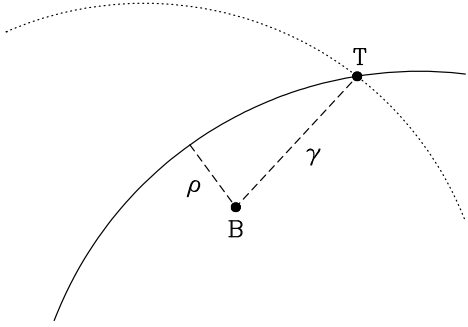,width=65mm}
\hspace{3mm}
\begin{minipage}[b]{90mm}
\caption{If two thin IPN annuli are available (solid and dotted arcs),
then we assume that their intersection gives the true location
$\hat {T}$.
In this case the angle $\gamma$ between the true location $\hat  {T}$
and the BATSE location $\hat  {B}$ is known.
If only one IPN annulus is available (solid arc), the position of
the true location on the annulus is unknown, and  the
separation between the annulus and BATSE location $\hat  {B}$
is characterized by the angle of closest approach $\rho$.}
\protect \vspace{3mm}
\end{minipage}
}
\end{center}
\end{figure*}

Our analysis approach assumes that the reference locations are effectively
points or lines
compared to the BATSE locations, an assumption which is fulfilled 
by eliminating all annuli with $3 \sigma$ widths exceeding $1.6^\circ$ and
eliminating $3 \sigma$ error boxes with any corner-to-corner dimension exceeding
$1.6^\circ$.
Because there are few IPN annuli with large widths, decreasing or increasing
the 1.6$^\circ$ requirement by a factor of two leads to only 
small changes in the
number of annuli included in the comparison sample and negligible changes in
the results.
We  convert the annuli to circles on the sphere, 
or their intersections to points,
by using the circle in the middle of each annulus.
Each pair of annuli has two intersections---we assume that the intersection
closest to the BATSE location is the true location.
Many of the intersecting  annuli create long error boxes which are
rejected by the requirement that no dimension exceed $1.6 ^\circ$,
in which case we keep only the narrowest annulus.
Such error boxes are common
during the time period of the 4Br catalog because frequently several of the
spacecraft of the Third Interplanetary Network have been close in space.
The vectors connecting each pair of spacecraft define the centers of the
annuli, and with some spacecraft in similar directions, these vectors
are typically close to parallel.
This leads to `grazing' intersections of the annuli and long error boxes.
(An additional  consequence of grazing intersections 
is that a small error in one annulus will move
the intersection a comparatively large distance).
After removing the wide annuli and large error boxes, 430 annuli for 411 GRBs
remain.   
For GRBs with intersecting annuli meeting our criteria
and therefore with a measurement of $\gamma$,
we do not use the two measurements of $\rho$ that could be obtained from
the individual annuli,
thereby ensuring the independence of all the $\gamma$ and $\rho$ measurements.
These criteria yield a
reference dataset with 19 measurements of $\gamma$ and 392 measurements
of $\rho$.

\begin{figure*}[tbp!]
\begin{center}
\mbox{
\psfig{figure=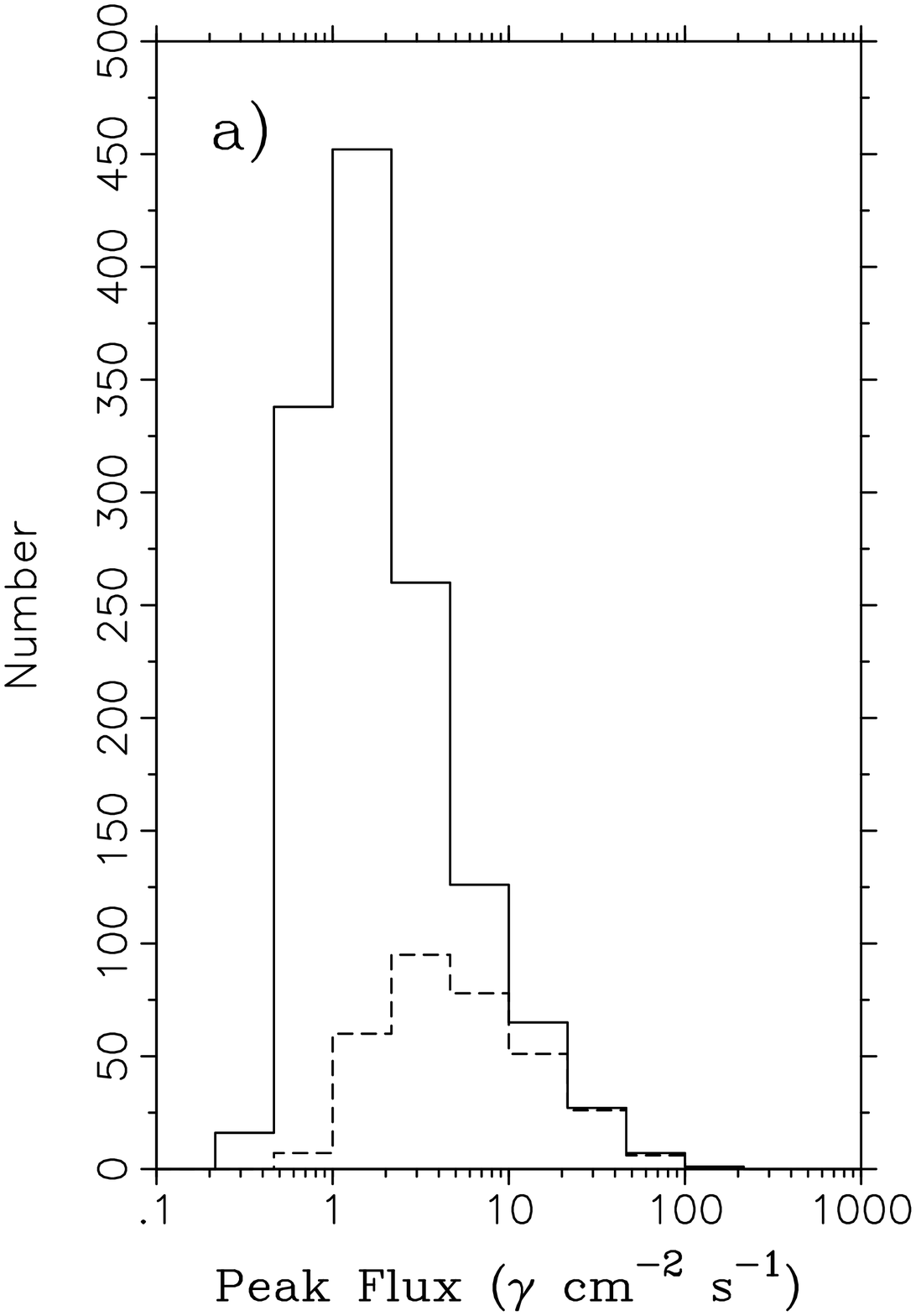,height=81mm}
\hspace{10mm}
\psfig{figure=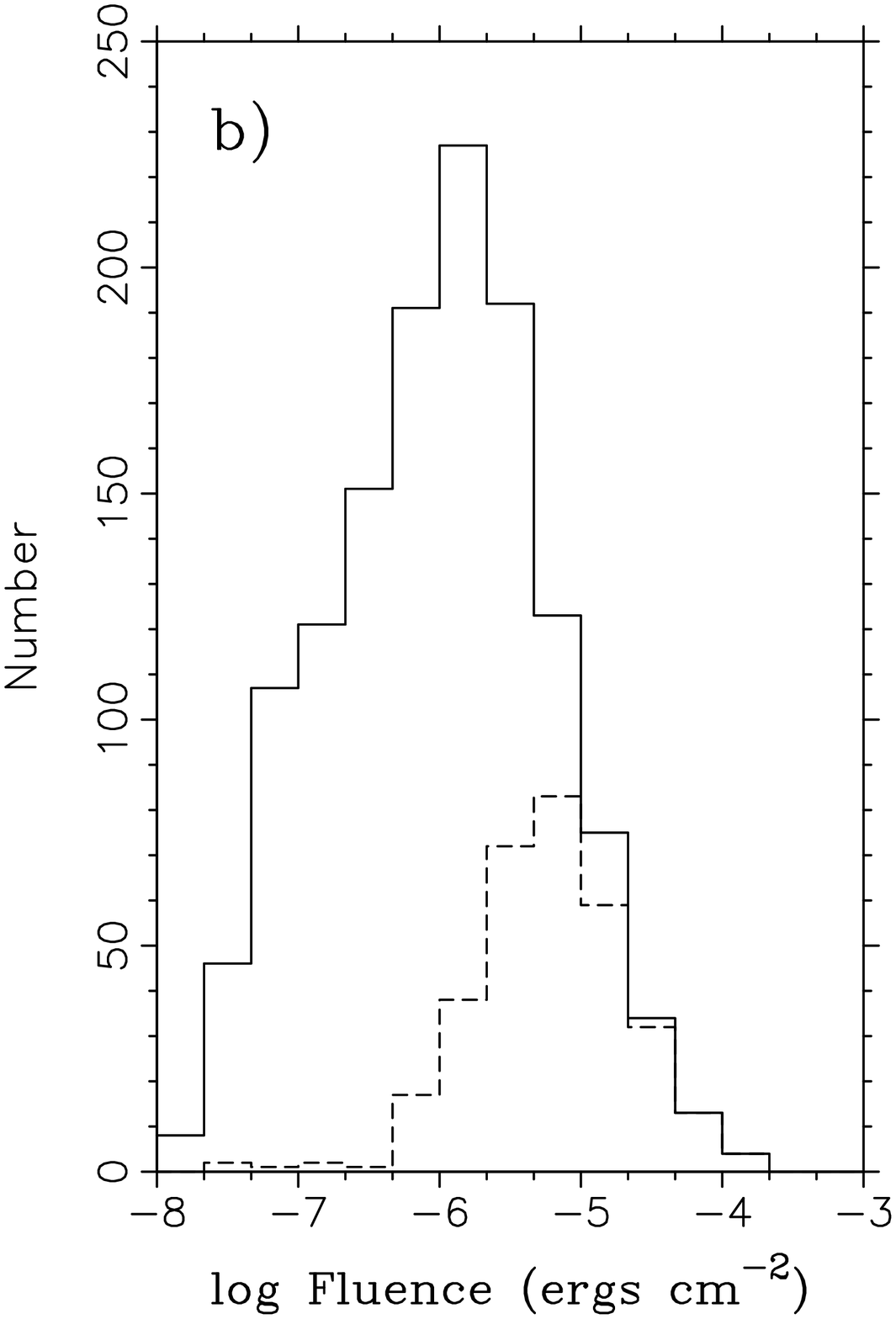,height=81mm}
}
\end{center}
\mbox{
\hspace{28mm}
\psfig{figure=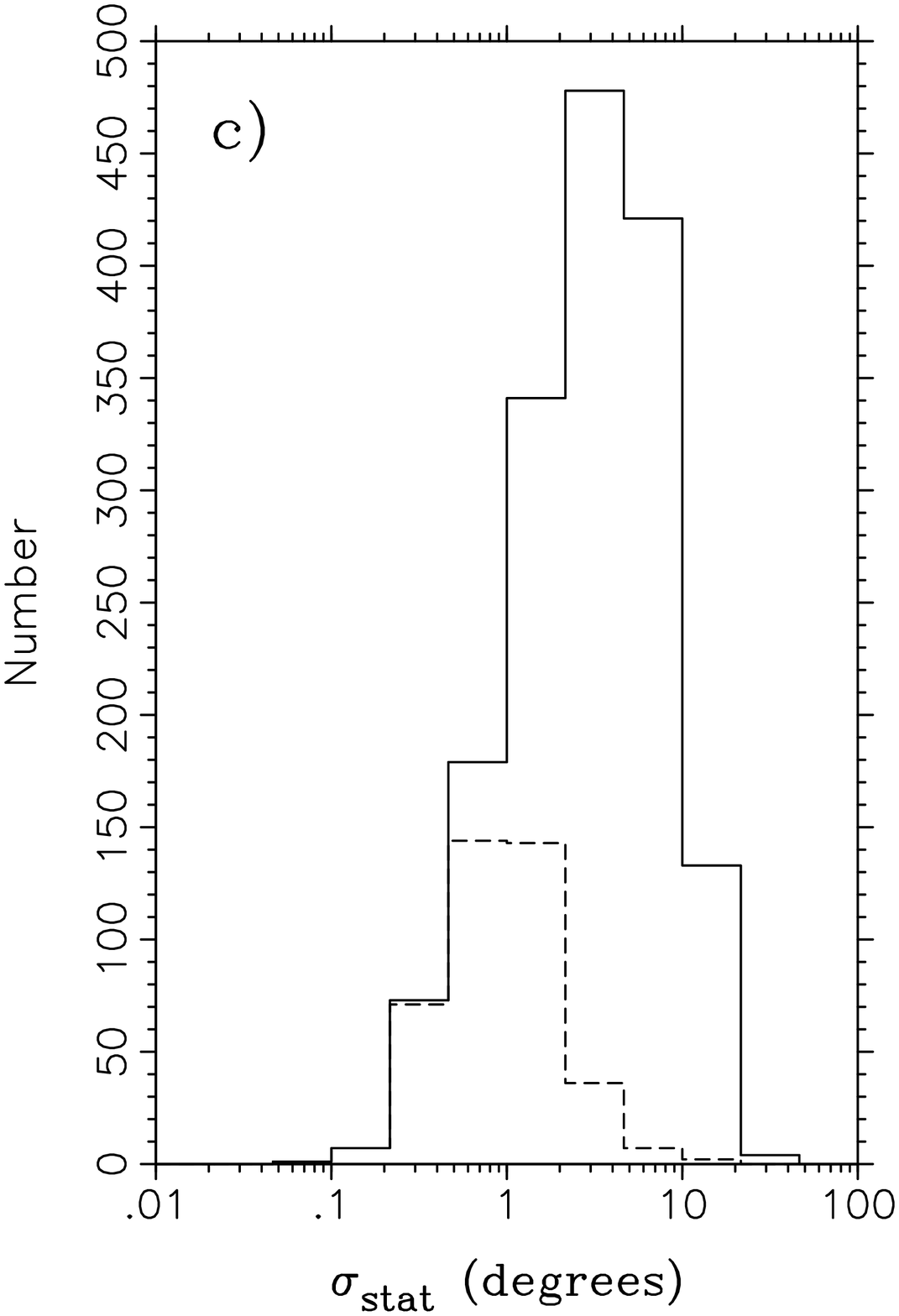,height=81mm}
}
\caption{Comparison of bursts for which BATSE triggered (solid
histograms) with the subset in the IPN Supplements 
(dashed histograms).
a) The distributions of peak flux as
observed by BATSE between 50 and 300 keV on the 64 ms timescale.
There are 1292 bursts
in the 4Br catalog with flux/fluence values (solid histogram), of which
324 are also in the IPN Supplements (dashed histogram).
b) The distributions of fluence as observed by BATSE between 50 and 300 keV
of the same groups of 1292 and 324 bursts, 
c) The distributions of BATSE $\sigma_{\rm stat}$ for all 1637 GRBs of the
4Br catalog (solid histogram) and for the subset of 411 events with
IPN data meeting our criteria (dashed histogram).
}
\end{figure*}

The 411 GRBs with IPN locations meeting our criteria represent 25\%
of the 1637 GRBs of the 4Br catalog.
The relative sensitivities of BATSE and the deep-space instruments of
the Third Interplanetary Network are largely determined by their effective
areas and by their backgrounds.    Each BATSE LAD has an effective area
of 2025 cm$^2$, compared to, for example, a projected area of
20~cm$^2$ for {\it Ulysses}.   
The differences in sensitivities are not as great as the comparisons of areas
would indicate
because the deep-space instruments have  lower and more stable
backgrounds.
Fig.~2 compares the properties of the bursts of the 4Br catalog with the
bursts of the IPN Supplements to the 4Br catalog.
The IPN Supplements sample almost the entire range of peak fluxes 
on the 64~ms timescale of the
4Br catalog, albeit with emphasis on the bright bursts.
In terms of fluence, the IPN Supplements have a strong bias towards bright
bursts, with only rare events below $10^{-6}$ erg cm$^{-2}$ having IPN
annuli.
There is a close correlation between burst fluence and BATSE
$\sigma_{\rm stat}$, so a similar effect is seen in the
distributions of $\sigma_{\rm stat}$.

\subsection{Bayesian Model Comparison}

If many `point' reference locations were available, a histogram
of the observed separations $\gamma$
between the BATSE locations and the reference
locations would provide a good representation of the probability distribution
of BATSE errors.   Because most of the reference locations from the IPN
Supplements are single annuli, insufficient information  exists for such
a simple approach.  Instead the error function
$p(\gamma)$ is determined by comparing models for $p(\gamma)$
and optimizing their parameters
using the entire dataset of $\gamma$ and $\rho$ measurements.

A Bayesian analysis very similar to that of
\markcite{GL96} Graziani and Lamb (1996) is used
to identify the best error model.
The Bayesian approach identifies the likelihood function as the
correct `merit' function \markcite{Siv96} (Sivia 1996).
The likelihood function $L$ is simply the product of the probability
density functions $p$ evaluated at each observed value of $\gamma$ for bursts
$i$ with intersecting annuli and at each observed value of $\rho$ for
bursts $j$ with single annuli:
        \begin{equation}
            L = \prod_{i} p_\gamma (\gamma_i)  \prod_{j} p_\rho (\rho_j).
        \end{equation}
The probability density function $p_\rho$ used for the single annuli cases
is analytically derived from $p_\gamma$ (see Appendix).
The likelihood function identified by the
Bayesian method specifies how  measurements of differing nature
(intersecting vs. single annuli) are to be combined in the analysis.
While eq.~1 shows the likelihood $L$ as a function of the observations, 
$L$ is also, 
through the functions $p_\gamma$ and $p_\rho$,
a function of the error model and its parameters.
For a particular error model, the best-fit parameters are obtained by
maximizing the likelihood.

An additional consideration arises in comparing  models: a model with
more parameters than another must be penalized because it has greater ability
to match the data whether or not it is correct.
The Bayesian approach   specifies how this must be done: each
model is penalized by an Occam's factor $F$.
Assuming that prior to examining the data
one expects parameter $\lambda_k$ to be in the range
$\lambda_k^{\rm min}$ to $\lambda_k^{\rm max}$ and that the likelihood
function is approximately Gaussian, these factors are, per parameter 
$\lambda_k$ \markcite{Siv96} (Sivia 1996),
\begin{equation}                            
F_k = \frac{ \sigma_{\lambda_k} \sqrt{2\pi} } 
{\lambda_k^{\rm max} - \lambda_k^{\rm min}},
\end{equation}
where $\sigma_{\lambda_k}$ is the uncertainty on $\lambda_k$ obtained
from the fit.
The overall Occam's factor $F$ is the product of the $F_k$ and 
penalizes a model for obtaining a better fit by using
adjustable parameters, with a greater penalty (smaller $F$) when the parameters
have larger possible prior ranges.

The final consideration is that one might have prior preferences for particular
models from earlier analyses or judgments .    These preferences can
be represented as prior probabilities for each model $P_{\rm prior}(M)$.
Combining these considerations, the Odds ratio $O_{B/A}$ specifies
the factor by which one should prefer model B over model A:
        \begin{equation}
              O_{\rm B/A} =  \frac {P({\rm B})} {P({\rm A})}
      =  \frac {P_{\rm prior}({\rm B}) \times L({\rm B}) \times F({\rm B})}
               {P_{\rm prior}({\rm A}) \times L({\rm A}) \times F({\rm A})}.
        \end{equation}
To avoid suspicions that our beliefs have created the results, we
equate all prior probabilities so that the first factor in eq.~3 
cancels.
Many of the models have common parameters so that some of the Occam's factors
$F_k$ also cancel.
For most of the model comparisons the likelihood ratios are so large that
no reasonable priors nor Occam's factors $F$ could alter which model is
identified as best.

Further discussion of Bayesian methodology, model fitting and comparison are
given by \markcite{Siv96} Sivia (1996) and \markcite{Lor90} Loredo (1990).

Instead of listing the odds ratios $O_{\rm B/A}$ for every combination
of models~A and~B, we list what we call odds factors $O_{\rm M}$ for
each model~M.   The odds factors $O_{\rm M}$ are proportional to
$P({\rm M})$ by a factor which is the same for each model, therefore the
odds ratios  $O_{\rm B/A}$ can be found by forming the ratio
of the factors $O_{\rm B}$ and
$O_{\rm A}$, or by differencing their logarithms (Tables 2 and 5).
The likelihoods and parameter uncertainties are also listed (Tables 2 and 3)
so that others can calculate odds ratios based upon their priors.

While the quantitative analysis is done with unbinned data using likelihood,
we prefer to depict the data and models with histograms (Fig. 3--5, 7 \& 9).
In order to present the results in a consistent manner, 
the data and models are binned using the total  errors of the minimal model,
$\sigma^0_{\rm tot}$ $=(\sigma_{\rm stat}^2 + 1.6^\circ \:^2)^{1/2}$.

\subsection{Probability Models}

We define the total error as the quadratic sum of 
the statistical uncertainty
$\sigma_{\rm stat}$ (as listed in the 4Br catalog)
and a systematic error:
\begin{equation}
\sigma_{\rm tot} = (\sigma_{\rm stat}^2 + \sigma_{\rm sys}^2)^{1/2}.
\end{equation}

Our error models are based on the
Fisher probability density function $p_{\rm F}$,
which has been called the Gaussian distribution on the sphere:
\begin{equation}
p_{\rm F}(\gamma) \; d \Omega =
\frac{\kappa}{2 \pi ({\rm e}^{\kappa} - {\rm e}^{-\kappa})}
      {\rm e}^{\kappa \cos{\gamma}} \; d \Omega,
\end{equation}
where $\gamma$ is the angle between the measured and true location
\markcite {FLE87} (Fisher et al. 1987).
The first convenience of this distribution is that it is normalized on
the sphere and that the `volume' element is clearly solid angle $d \Omega$.
The distribution also has convenient analytic properties;
in the small angle approximation it reduces to the Gaussian distribution.
An algorithm to simulate random locations from a Fisher distribution
is given by \markcite{FLE87} Fisher et al. (1987).

The probability $P$ of the true location lying in a region of solid angle
$\Omega$ is found by integrating the probability density function $p$:
\begin{equation}                   
P = 
\frac {\kappa} {2 \pi ({\rm e}^{\kappa} - {\rm e}^{-\kappa})}
   \int_{\Omega} d\Omega \: {\rm e}^{\kappa \cos \gamma}.
\end{equation}
Specializing to the probability of the true location being in the ring
$\gamma_1 \leq \gamma \leq \gamma_2$, eq.~6 becomes
\begin{equation}
P =  \frac {\kappa} {{\rm e}^{\kappa} - {\rm e}^{-\kappa}}
\int_{\gamma_1}^{\gamma_2} d\gamma \sin \gamma \:
               {\rm e}^{\kappa \cos \gamma}.
\end{equation}

The parameter $\kappa$ is termed the concentration parameter.
BATSE GRB location errors have traditionally been specified in terms of
$\sigma$, which has been defined by the BATSE team as the radius of the
circle with the same area as the
68\% confidence ellipse.
Assuming the error box to be circular, and
setting $P=0.68$, $\gamma_1=0$ and $\gamma_2= \sigma_{\rm tot}$ in eq.~7,
one obtains, for $\sigma_{\rm tot}$ in 
{\it radians} and for $\sigma_{\rm tot}$ small ($\lesssim 20^\circ$):
\begin{equation}
\kappa = \frac {1} {(0.66 \sigma_{\rm tot})^2}.
\end{equation}

Models more complicated than a Fisher distribution are tested.
The first method of building more complicated models is to make
$\sigma_{\rm sys}$ a function of either
intrinsic burst properties or of instrumental factors which might
influence the quality of locations.
The second way in which more complicated models are formed
is to sum two Fisher distributions with different values of
$\sigma_{\rm sys}$:
\begin{equation}
p = f_1 p_{\rm F}^1 + (1-f_1) p_{\rm F}^2.
\end{equation}
In this case, 
for each burst, the 
values of the two total location errors $\sigma_{{\rm tot}, i}$ are
calculated from the statistical uncertainties
$\sigma_{\rm stat}$  listed in the 
catalog and the model values $\sigma_{{\rm sys},i}$ (for $i=1,2$):
\begin{equation}
   \sigma_{{\rm tot},i} =
     (\sigma_{\rm stat}^2 + \sigma_{{\rm sys},i}^2)^{1/2}.
\end{equation}

For the single annulus cases, the probability density function
$p(\rho)$ is analytically derived from the probability density function
$p(\gamma)$ (see Appendix).  Since $\rho \leq \gamma$
(see Fig.~1), $P(\rho < X) > P(\gamma <X)$.
Table~1 gives some cumulative probabilities $P$ 
for both $p_{\rm F}(\gamma)$ and
the corresponding $p(\rho)$.

\begin{deluxetable}{crr}
\tablecolumns{3}
\tablenum{1}
\tablecaption{Cumulative Values for the Distributions
of $\gamma$ and $\rho$ assuming the Fisher Distribution}
\tablehead{
    Angle $X$   &  \colhead{$P_{\rm F}(\gamma \leq X)$}    &  
       \colhead{$P(\rho \leq X)$}     }
\startdata
  $1 \sigma$  &   0.6826  &  0.8703  \nl
  $2 \sigma$  &   0.9898  &  0.9976  \nl
  $3 \sigma$  &   $>$0.9999  &  $>$0.9999  \nl
\enddata
\end{deluxetable}

\section{Results}

\subsection{4Br Catalog}

The systematic error of $1.6^\circ$ used in the minimal model
was determined using 50 comparison locations, some of which had
errors not much smaller than the BATSE errors 
\markcite{Mee96} (Meegan et al. 1996).
Fig.~3 compares the 4Br data to the minimal model---with the large and 
accurate IPN reference dataset, the minimal model is clearly seen
to be a poor fit to the data.
Model~1 generalizes the minimal model by optimizing the value of
the systematic error, obtaining $2.8^\circ$.
While a major improvement over the minimal model,
this model fits the data poorly (Fig.~4) and has
a much smaller odds factor than some other models (Table~2).

\begin{figure*}[tbp!]
\begin{center}
\mbox{
\hspace{12mm}
\psfig{figure=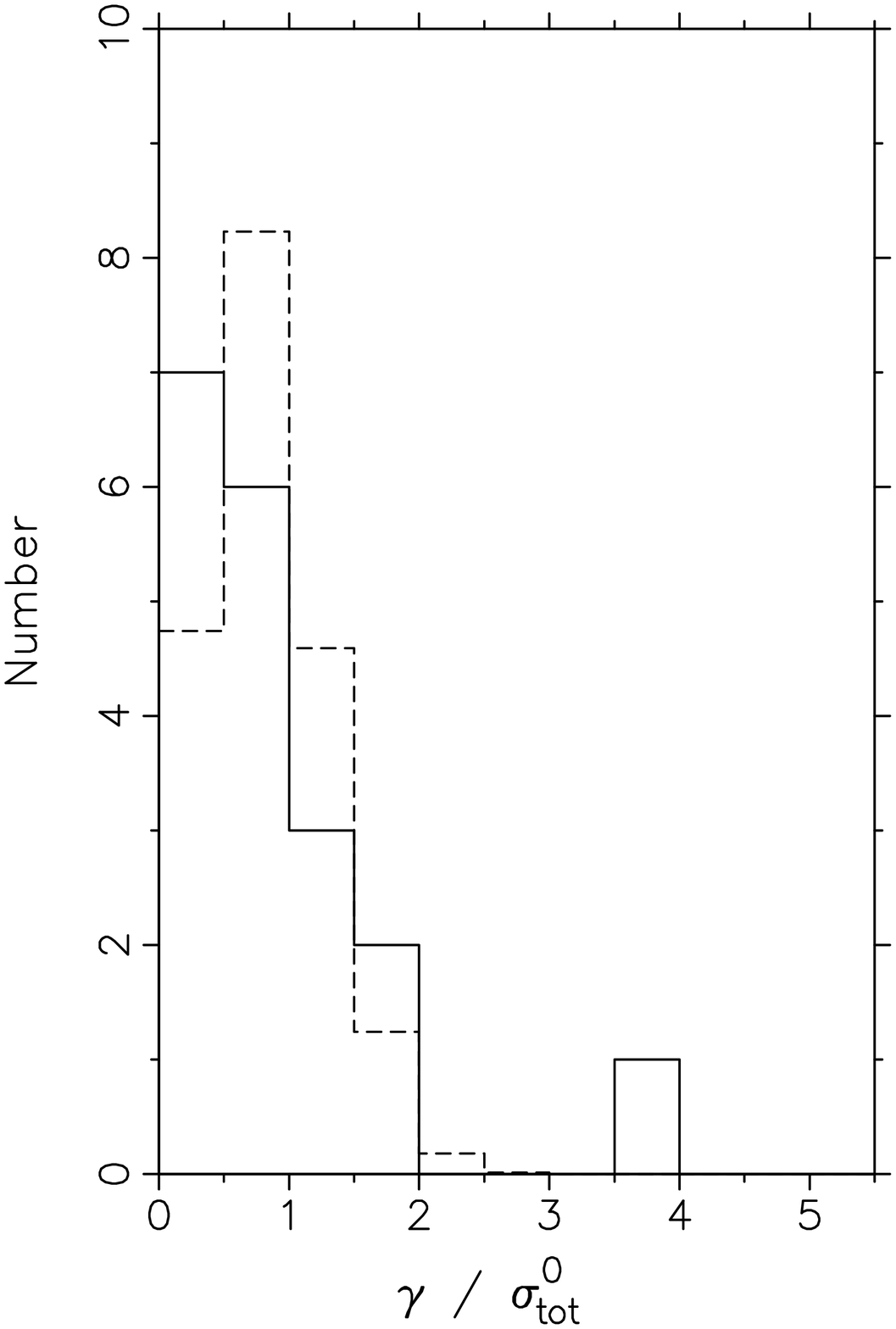,width=55mm}
\hspace{16mm}
\psfig{figure=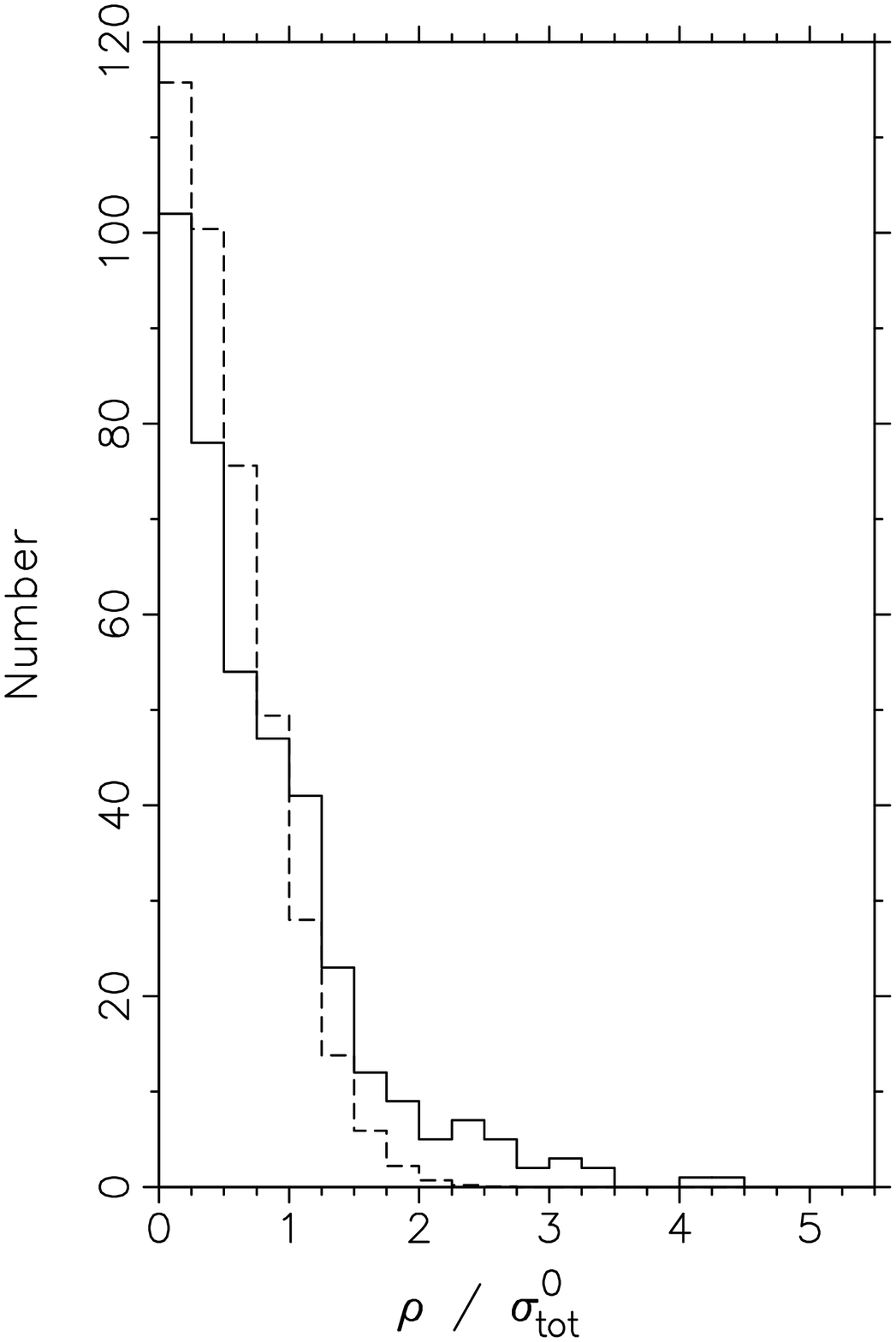,width=55mm}
\hspace{12mm}
}
\end{center}
\caption{Histograms of the data (solid) and Model~0 (dashed) for the
separations $\gamma$ between BATSE locations and intersecting IPN annuli
and the closest separations $\rho$ between BATSE locations
and single IPN annuli.
The data and model are binned in terms of $\sigma^0_{\rm tot}$
$= (\sigma_{\rm stat}^2 + 1.6^\circ \:^2)^{1/2}$.
The data and Model~0 are clearly in poor agreement.
}
\end{figure*}

The value of $1.6^\circ$ published in the 3B catalog 
\markcite{Mee96} (Meegan et al. 1996)
was determined by a
root-mean-square comparison of the observed separations with the combination of
the BATSE statistical uncertainties and the
known errors of the
reference locations.  
The 3B catalog warned that some locations might
be substantially worse than average and that the error distribution might
have a non-Gaussian tail.

\begin{figure*}[tbp!]
\begin{center}
\mbox{
\hspace{12mm}
\psfig{figure=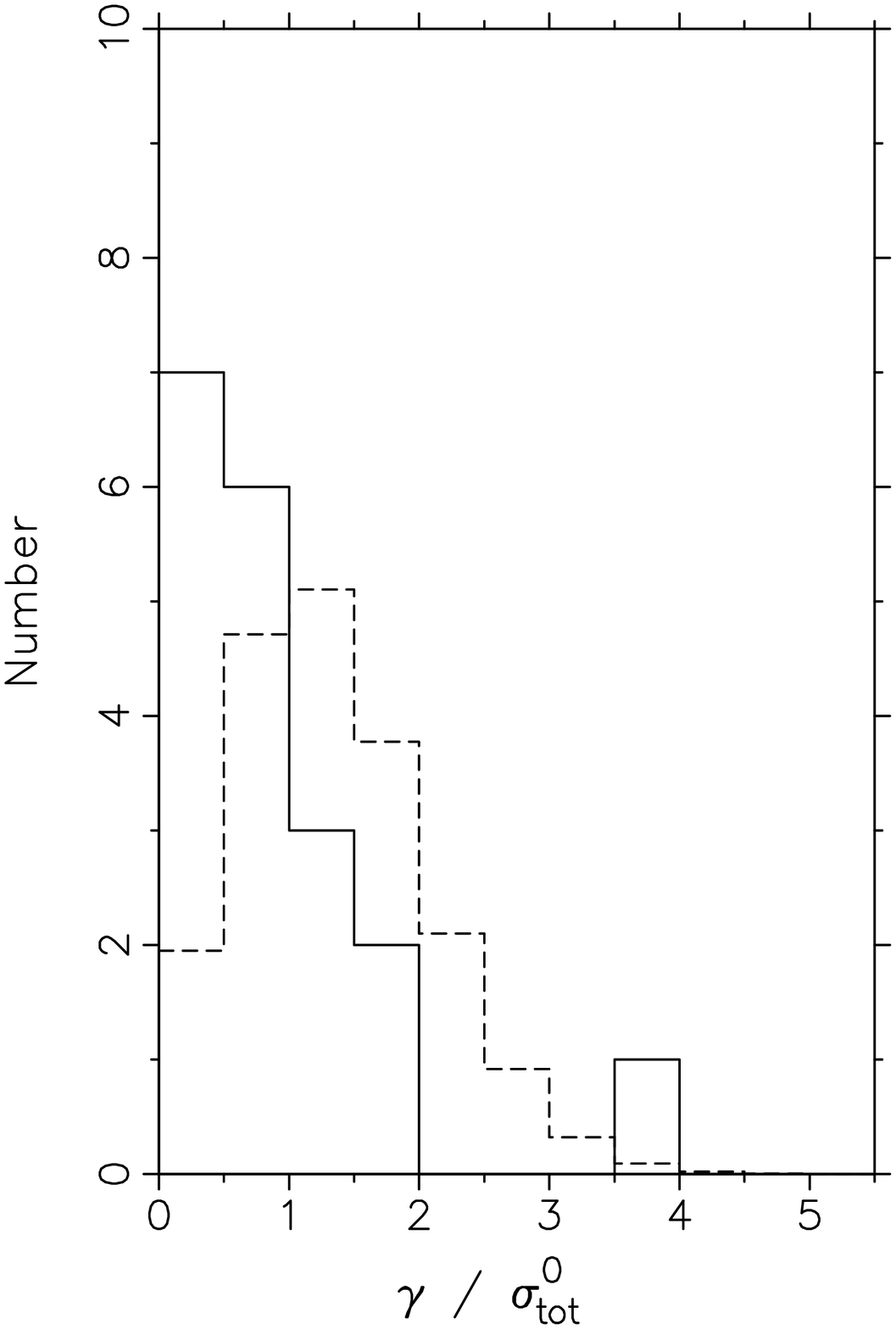,width=55mm}
\hspace{16mm}
\psfig{figure=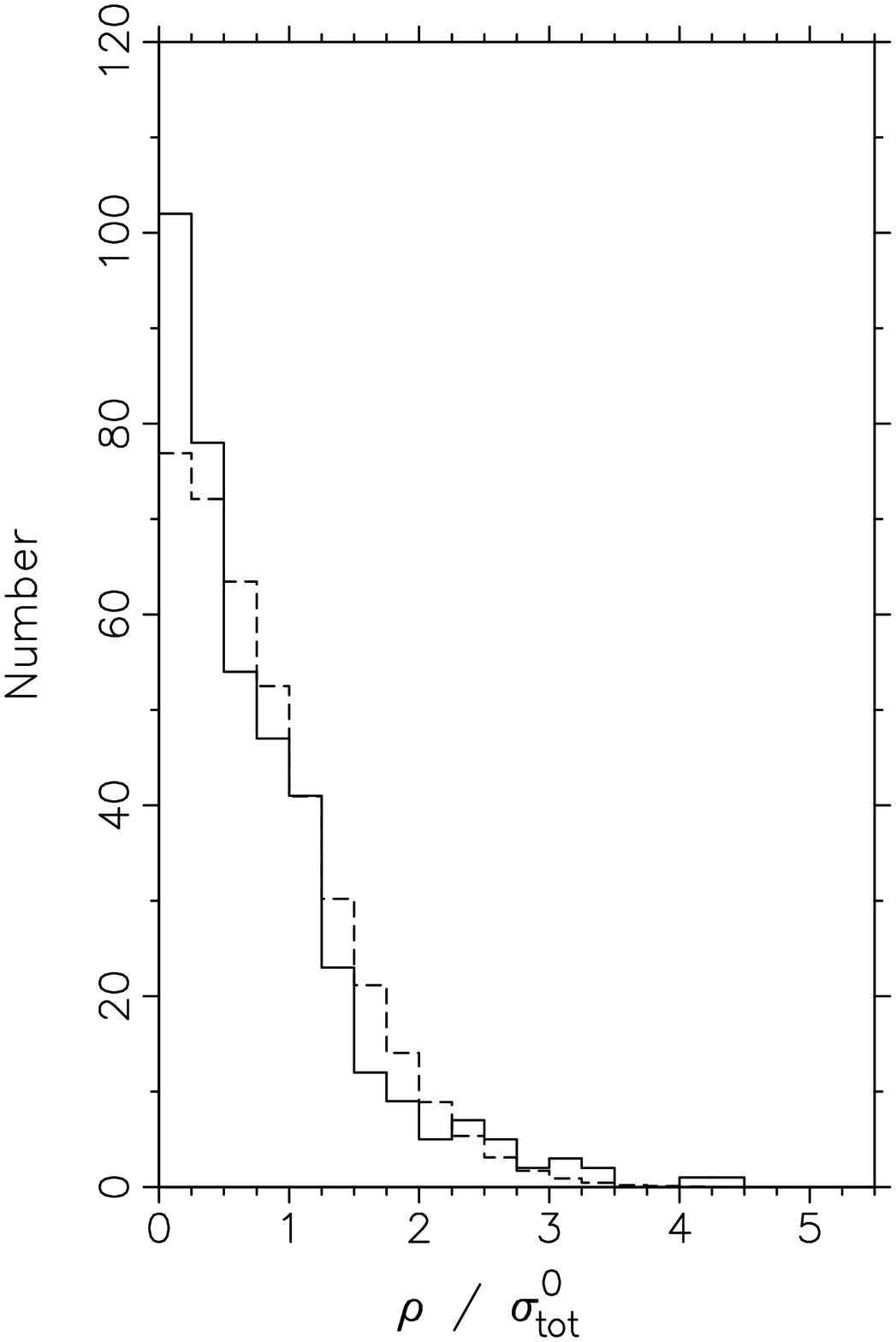,width=55mm}
\hspace{12mm}
}
\end{center}
\caption{Histograms of the data (solid) and Model~1 (dashed).
Model~1 predicts too few events at both small
($\rho / \sigma_{\rm tot}^0 \lesssim 0.25$) and large
($\rho / \sigma_{\rm tot}^0 \gtrsim 3$) separations.
}
\end{figure*}

Model~2 implements the idea of a Gaussian-like distribution with an
extended tail by  summing two Fisher distributions having
differing systematic errors (eqs.~9 and~10).
This model provides an excellent fit to the data (Fig.~5 and Table~2).
The evidence favors Model~2 over the minimal model by an odds ratio of
$10^{56}$ and over Model~1 by $5 \times 10^{7}$ (Table~2).
The best-fit parameters are given in Table~3.

\begin{figure*}[tbp!]
\begin{center}
\mbox{
\hspace{12mm}
\psfig{figure=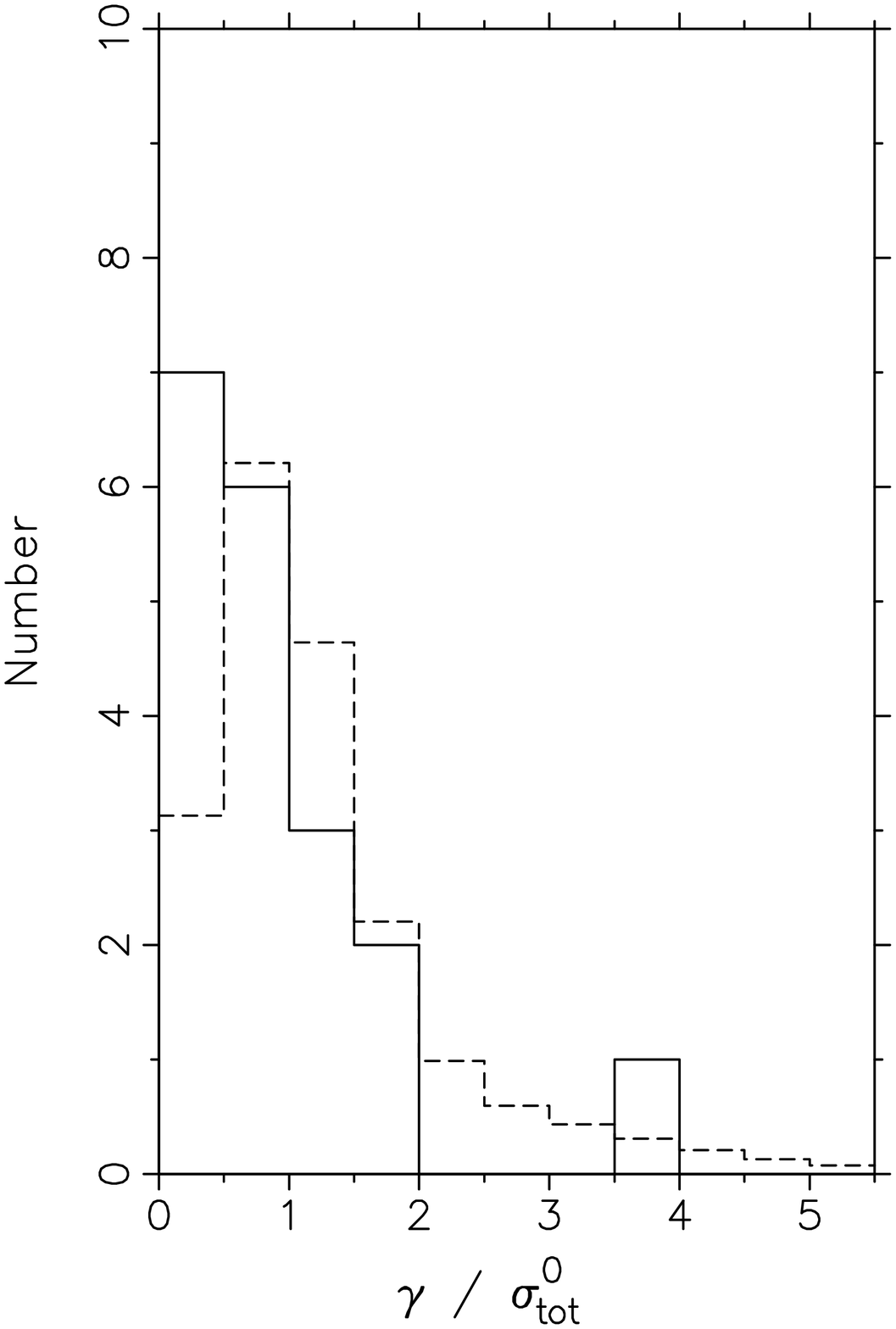,width=55mm}
\hspace{16mm}
\psfig{figure=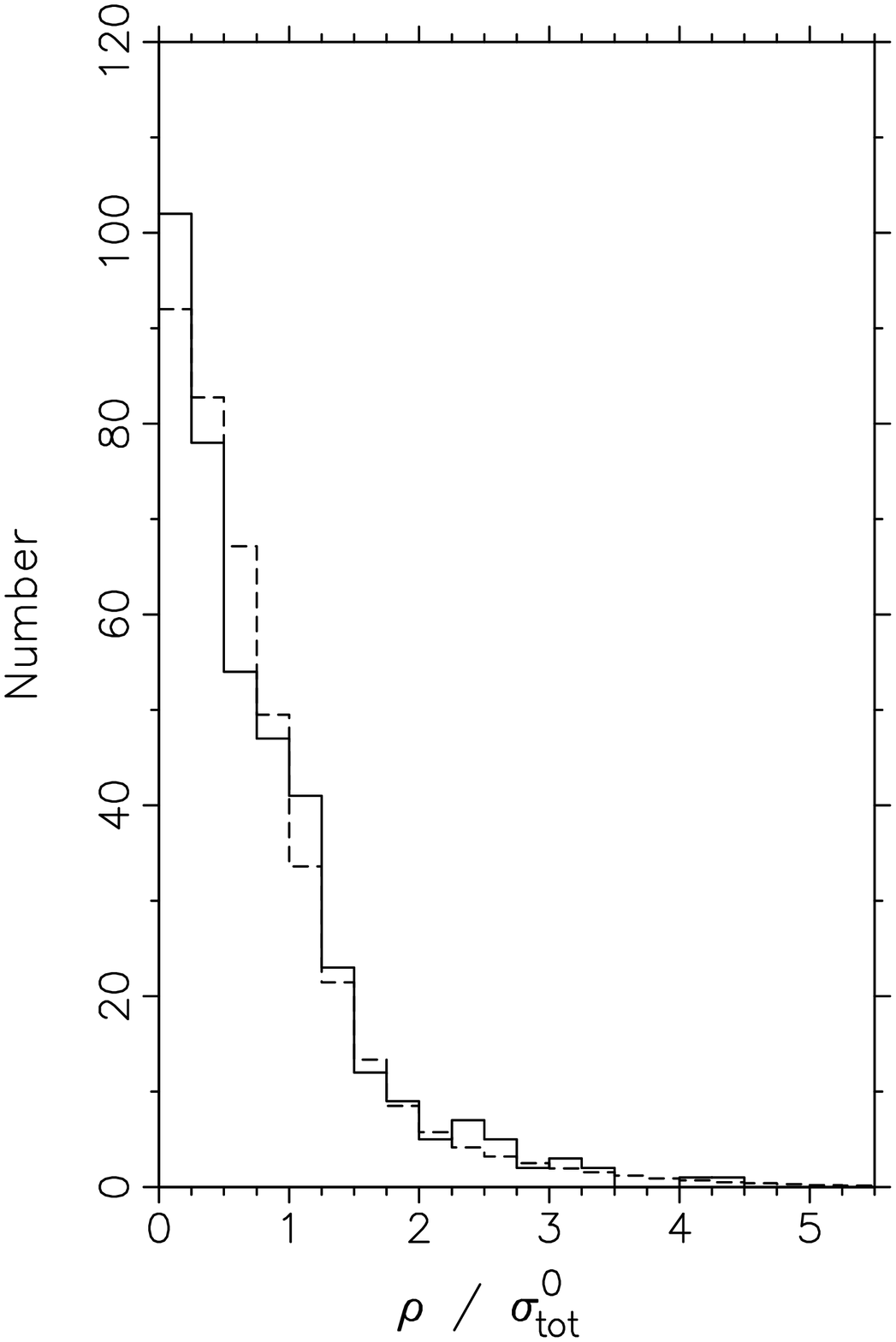,width=55mm}
\hspace{12mm}
}
\end{center}
\caption{Histograms of the data (solid) and Model~2 (dashed).
The agreement between the data and Model~2 is excellent.
}
\end{figure*}

\begin{deluxetable}{ccccc}
\tablecolumns{5}
\tablenum{2} 
\tablecaption{Model Comparison Results for 4Br Catalog}
\tablehead{
  &   &    Number of  &  $\log_{10}$   &   $\log_{10}$  \nl
  Model  &   Description    &  
     Parameters  &  likelihood &  Odds Factor }
\startdata
   0    &
\begin{minipage}[t]{45mm}
\begin{flushleft}
Minimal: Systematic Error
Assumed to be $1.6^\circ$
\end{flushleft}
\end{minipage}  &
 0  &  161.5  &  161.5  \nl
\tablevspace{3mm}
   1    & 
\begin{minipage}[t]{45mm}
\begin{flushleft}
Single Systematic Error applied to all locations
\end{flushleft}
\end{minipage} &
1 &  211.5  &  210.2  \nl
\tablevspace{3mm}
    2   &
\begin{minipage}[t]{45mm}
\begin{flushleft}
Core-plus-Tail Systematic Errors
\end{flushleft}
\end{minipage} &
3  &    220.8  &  217.9  \nl
\tablevspace{3mm}
   4   &
\begin{minipage}[t]{45mm}
\begin{flushleft}
Systematic Error a power law
function of statistical uncertainty:
$\sigma_{\rm sys} = A (\sigma_{\rm stat} / 1^\circ)^\alpha$
\end{flushleft}
\end{minipage}    &
2  &  212.6  &  210.4  \nl
\tablevspace{3mm}
    10   &
\begin{minipage}[t]{45mm}
\begin{flushleft}
Core-plus-tail Systematic Errors depending on datatype
\end{flushleft}
\end{minipage} &
 6  &  223.7  &  219.4  \nl
\tablevspace{3mm}
   12  &
\begin{minipage}[t]{45mm}
\begin{flushleft}
Core-plus-tail for CONT, one systematic error for other datatypes
\end{flushleft}
\end{minipage}  &
 4  &  222.9  &  219.0  \nl
\tablevspace{3mm}
   33  &
\begin{minipage}[t]{45mm}
\begin{flushleft}
Core-plus-tail, only core systematic error depends on datatype
\end{flushleft}
\end{minipage}  &
 4  &  223.6  &  219.9  \nl
\enddata
\end{deluxetable}

\begin{deluxetable}{cccc}
\tablecolumns{4}
\tablenum{3}
\tablecaption{Best-Fit Error Model Parameter Values for the 4Br Catalog}
\tablehead{
 Model  &   \multicolumn{3}{c}{Parameters}  \nl
    &   Symbol  &   Description  &   Value\tablenotemark{\dagger}  }
\startdata
   1    &   $\sigma_{\rm sys}$  &
  Systematic Error  &   $2.80 \pm 0.12$  \nl
\tablevspace{3mm}
    2   &   $\sigma^1_{\rm sys}$  &  Core systematic error & 
   $1.85^\circ \pm 0.16^\circ$ \nl
            &   $f_1$  &  Fraction in Core Term  &  $0.78^ \pm 0.08$  \nl
            &   $\sigma^2_{\rm sys}$  &  Tail systematic error & 
   $5.1^\circ \: ^{+0.8}_{-0.6} $ \nl
\tablevspace{3mm}
   4  &  $A$  & Amplitude (error for $\sigma_{\rm stat} = 1^\circ$)  &
   $2.93 \pm 0.13$  \nl
      &  $\alpha$  &  Power law index  &
   $0.14 \pm 0.06$  \nl
\tablevspace{3mm}
    10   &   $\sigma_{\rm CONT}^1$  &  Core systematic error for CONT & 
      $1.68^\circ \pm 0.16^\circ$ \nl
       &   $f_{\rm CONT}^1$  &  
Fraction in Core Term for CONT &  $0.82 ^{+0.05}_{-0.08}$  \nl
         &      $\sigma_{\rm CONT}^2$  &  Tail systematic error for CONT & 
   $5.3^\circ \: ^{+1.0}_{-0.8} $ \nl
  &  $\sigma_{\rm other}^1$  &  Core systematic error for non-CONT & 
   $2.6^\circ \pm 0.5^\circ$ \nl
      &   $f_{\rm other}^1$  &  
Fraction in Core Term for non-CONT &  $0.73 ^{+0.19}_{-0.35}$  \nl
      &     $\sigma_{\rm other}^2$  &  Tail systematic error for non-CONT & 
   $5.2^\circ \: ^{+2.1}_{-1.2} $   \nl
\tablevspace{3mm}
    12   &   $\sigma_{\rm CONT}^1$  &  Core systematic error for CONT & 
   $1.67^\circ \pm 0.15^\circ$ \nl
       &   $f_{\rm CONT}^1$  &  
Fraction in Core Term for CONT &  $0.82 ^{+0.05}_{-0.08}$  \nl
         &      $\sigma_{\rm CONT}^2$  &  Tail systematic error for CONT & 
   $5.3^\circ \: ^{+1.0}_{-0.8} $ \nl
  &  $\sigma_{\rm other}^1$  &  Systematic error for non-CONT & 
   $3.44^\circ \pm 0.26^\circ$ \nl
\tablevspace{3mm}
    33   &   $\sigma_{\rm CONT}^1$  &  Core systematic error for CONT & 
   $1.67^\circ \pm 0.15^\circ$ \nl
         &   $\sigma_{\rm other}^1$  &  Core systematic error for other & 
   $2.74^\circ \pm 0.34^\circ$ \nl
       &   $f_{\rm all}^1$  &  Fraction in Core, all datatypes & 
   $0.82  ^{+0.05}_{-0.07} $  \nl\
         &      $\sigma_{\rm all}^2$  &  
   Tail systematic error for all  datatypes & 
   $5.4^\circ \pm 0.8^\circ $ \nl
\enddata
\tablenotetext{\dagger}{The uncertainties on the parameter values are for
single parameters of interest (i.e., they were obtained from the
change in likelihood equivalent to the
usual $\chi^2 +1$
prescription).   In some cases the uncertainties between
the parameters are highly correlated.  However, further specification of the
errors on the errors of the locations would be excessive.}
\end{deluxetable}

Model 2 is
best thought of as a `core-plus-tail' representation of the
error distribution rather than as a two-component model.
The model does not specify
that some bursts 
belong to a core component (small $\sigma_{\rm sys}$) and others to a
tail component (large $\sigma_{\rm sys}$);
instead all bursts have a tail to their location error 
distribution.  Model~2 uses no burst property other than $\sigma_{\rm stat}$
to determine the location error distribution of each burst.

To depict the distribution of separations $\gamma$ between the BATSE location
and the true location given by Model~2, we perform part of the solid angle
integration over the sphere of the probability density function
$p_2$ of Model~2:
\begin{equation}
1 = \int_{0}^{\pi} d\gamma \: \int_0^{2 \pi}  d\psi \: \sin \psi \: p_2(\gamma)
\end{equation}
\begin{equation}
  = \int_{0}^{\pi} d\gamma \: \frac{dP_2}{d\gamma}(\gamma) .
\end{equation} 
The function $dP_2 / d\gamma $ is depicted in Fig.~6 for
a burst with a negligible value for $\sigma_{\rm stat}$.
The `core' and `tail' terms are shown separately along with the total model.
While 22\% of the probability is in the tail term, much of the
area of the tail term is near $\gamma=0$ so that only 7\% of the
locations are past $\gamma=5^\circ$.

\begin{figure*}[tbp!]
\begin{center}
\mbox{
\psfig{figure=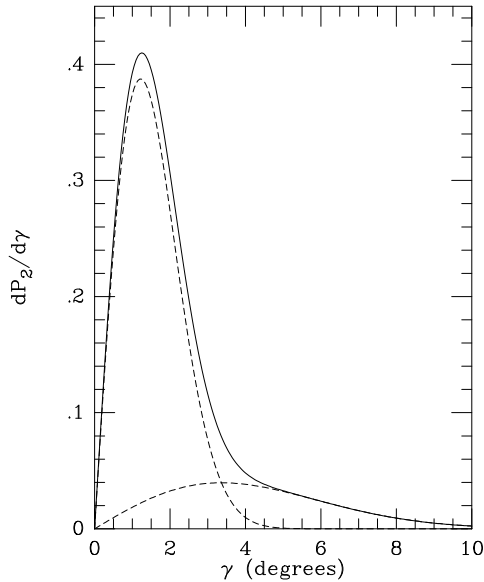,width=65mm}
\hspace{3mm}
\begin{minipage}[b]{90mm}
\caption{The function $d P_2 / d\gamma$ (eqs.~11 and 12) is
the error distribution versus separation $\gamma$ according to
Model~2.
It is shown 
for a gamma-ray burst with a negligible value for $\sigma_{\rm stat}$.
The function as depicted includes the solid
angle factor $2 \pi \sin \psi$.  If this factor were not included, the 
function
would be very large near $\gamma=0$, yet the probability of very small $\gamma$
is low because of the solid angle factor.
The integral from $\gamma= 0$ to $180^\circ$ 
of the function shown is one.
The solid curve shows Model~2 (as specified in Table~3), while the dashed
curves show the two terms, `core' and `tail', separately.
}
\protect \vspace{14mm}
\end{minipage}
}
\end{center}
\end{figure*}

We have endeavored to find correlations that would allow us to assign
$\sigma_{\rm sys}$ values based upon burst or location properties, either
by fitting models with $\sigma_{\rm sys}$ a function of the property, or
by dividing the locations into categories by the property and comparing
the values of model parameters.
Among the properties for which no significant dependence was found are:
the fluence of the burst,
the value of $\sigma_{\rm stat}$,
the $\chi^2$-value obtained by LOCBURST, 
the angle to the center of the earth (which affects the importance
of atmospheric scattering),
the spectral index of the burst,
and whether the event was a {\it Ulysses} trigger or was obtained
from {\it Ulysses} continuous data.
In a test for a dependence of location quality on spacecraft scattering,
we made three extensions each of Models 1 and 2, adding a parameter dependence
on the spacecraft hemisphere ($+/-$X, $+/-$Y, or $+/-$Z) in which the burst was
located.   The coordinate system is defined so that COMPTEL and EGRET view in
the +Z direction and OSSE scans in the XZ plane 
\markcite{Geh93} (Gehrels, Chipman \& Kniffen 1993).
The motivation is that the spacecraft is not symmetrical in Y and Z, and
inadequacies in the modeling of the scattering from the spacecraft might cause
location quality to depend on position in spacecraft coordinates.
There were modest odds ratios improvements
for two of the extensions of
Model~1 (not approaching the odds ratio of Model 2), but none for the
extensions of Model~2.   We conclude that the bilateral asymmetries of the
spacecraft do not create bilateral asymmetries in location quality.

The only property that we have  identified as significant
is the type of the data used to obtain the location.
In addition to four
continuously transmitted `background' datatypes, the
BATSE instrument  packages data from triggered events into
eight burst datatypes.    
Any of the background or burst datatypes which
include data separately from each of the LADs can be used to derive
a location.
The default datatype for locating bursts is CONT, which provides continuous
coverage regardless of trigger status
in 16 energy channels with 2~s time resolution.
The CONT datatype was used for 58\% of the locations in the 4Br catalog.
If the event is shorter than 2~s, so that using CONT data would 
unnecessarily add statistical fluctuations, or if CONT data are
unavailable due to a telemetry gap, discriminator data 
(DISCLA, DISCLB, PREB or TTE)  with four energy channels are used.
If normal telemetry is available, 1~s resolution
DISCLA would typically be used.
A duplicate of DISCLA data, DISCLB, is included in the burst data
to obtain more reliable telemetry.
Together DISCLA and DISCLB data were used for 23\% of the locations.
The instrument triggers at the end of the interval in which a significant
rate increase is registered.   If the burst has essentially ended by the
trigger time, then one of the high-time resolution pre-burst datatypes is
used, either PREB or TTE.    This was the case for 12\% of the locations.
Finally, if no regular data are available due to missing telemetry, the
maximum rates recorded by the instrument during the burst readout are used
(MAXBC or MAXC1).
This was necessary for 8\% of the 4B locations.   (The percentages sum to
101\% due to rounding.)

Models 10, 12 (\S3.2) and 33 incorporate
correlations between location accuracy and datatype.
Model~10 is very similar to Model~2, with the difference that the
model parameter values depend on the datatype.
One set of model parameters like those of Model~2 are
used for locations obtained using CONT data,
$\sigma^1_{\rm CONT}$, $f^1_{\rm CONT}$, and $\sigma^2_{\rm CONT}$,
while another set,
$\sigma^1_{\rm other}$, $f^1_{\rm other}$, and $\sigma^2_{\rm other}$,
is used for all other locations.
Model~10 has a likelihood $\times 800$ larger than that of Model~2 (Table~2).
Considering the three additional parameters, the odds ratio favors
Model~10 by a factor of 30.

Two pairs of parameters of Model 10,
$f^1_{\rm CONT}$ \&  $f^1_{\rm other}$ and
$\sigma^2_{\rm CONT}$ \& $\sigma^2_{\rm other}$, have values which are
consistent (see Table~3), i.e.,
the data do not demonstrate that these parameters have
different values.    This suggests a model with fewer parameters:
Model~33 has distinct parameter values for the core systematic errors
of CONT-derived locations and locations obtained with other data types,
but only a single parameter $f^1_{\rm all}$ representing the probability
of the core term, and only a single tail
systematic error $\sigma^2_{\rm all}$.
The likelihood ratios of Models~10 and~33 are virtually
identical; because of the fewer
parameters of Model~33 its odds ratio is $\times 3$ better (Table~2).
This improvement may mean that the `tail' is caused by factors independent
of datatype, or simply that
there is insufficient data to discern differences in
the tail as a function of datatype.   The odds ratio improvement is marginal
and the models make very similar predictions, so an explanation is not
required.
Model~33 is favored over Model~2 by an odds ratio of 100.
Qualitatively, this might be termed as  
persuasive but not compelling evidence for  Model~33 over Model~2. 
The data and Model~33 are compared in Fig.~7.

\begin{figure*}[tbp!]
\begin{center}
\mbox{
\hspace{12mm}
\psfig{figure=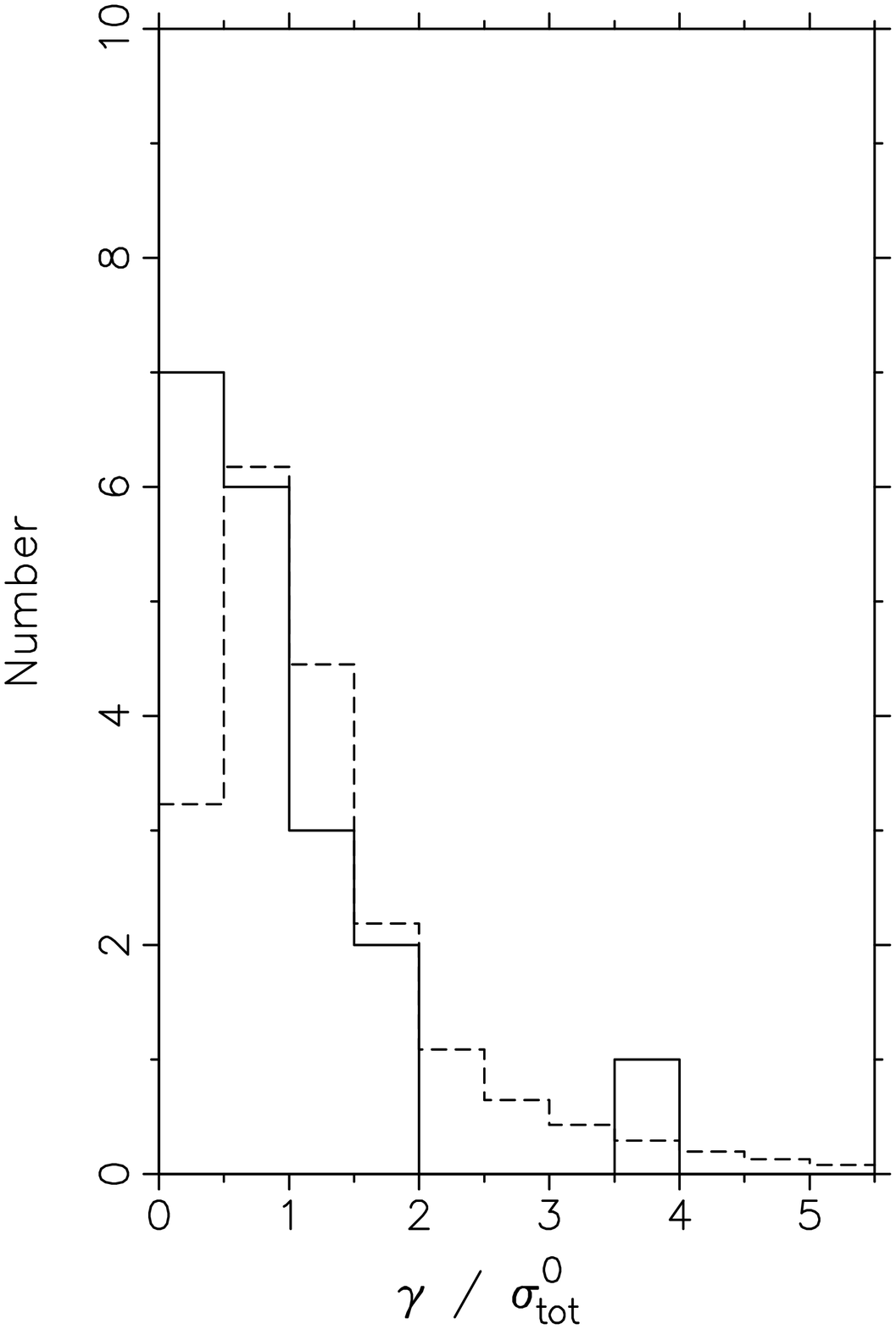,width=55mm}
\hspace{16mm}
\psfig{figure=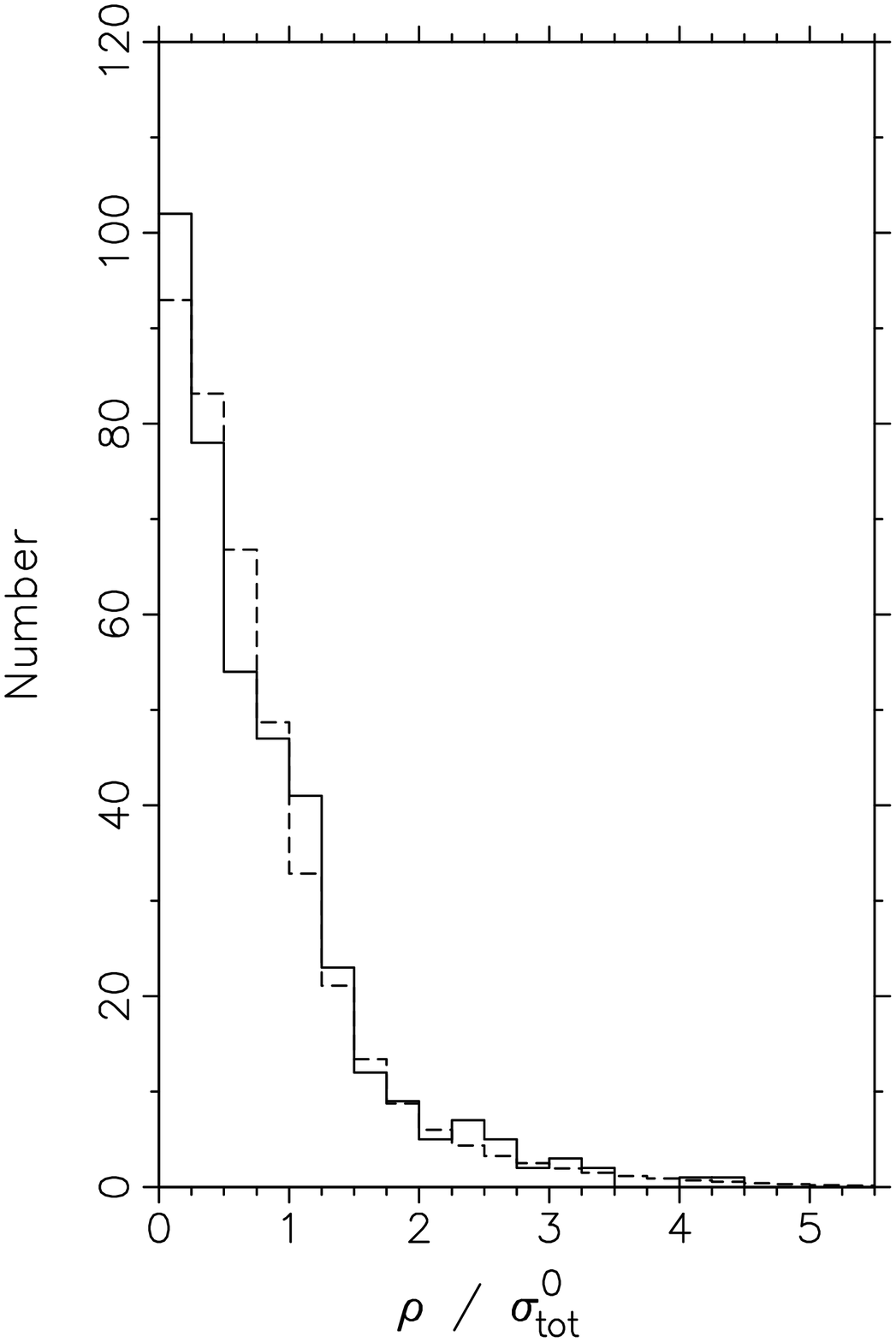,width=55mm}
\hspace{12mm}
}
\end{center}
\caption{Histograms of the data (solid) and Model~33 (dashed).
While quantitative measures show Model~33 to be better than Model~2,
the model histograms are almost identical.
}
\end{figure*}

Locations obtained with datatypes other than CONT, i.e., using either
PREB, TTE, DISCLA, DISCLB, or MAXBC datatypes, are consistent with
being of equal quality.
However, of the 411 bursts with reference IPN locations, only 110 were
located with datatypes other than CONT, so the comparisons are probably
insensitive.

The Bayesian approach relies upon model comparisons rather than goodness-of-fit
tests---generally one does not know whether a particular best-fit value of
the likelihood is reasonable.    As a final test of the quality of the models,
we (not being doctrinaire Bayesians)
perform goodness-of-fit tests via simulations.   Assuming a specific model
to be true, we simulate datasets of BATSE
locations using the best-fit parameters of the model.
We use the actual IPN annuli, pick locations on  single annuli
and on the intersection of each  pair to be the 
`true' locations, and then create simulated  BATSE locations using
the assumed error model.
The separations between the annuli and the simulated locations yield
simulated values of $\rho$ and $\gamma$.
The simulated dataset is fit with the assumed model, re-optimizing
its parameters and obtaining the best-fit value of the likelihood.
This process is repeated to generate 1000 simulated datasets and likelihood
values.
If the actual best-fit likelihood value is much different from  the
simulated best-fit likelihoods, the model is probably a poor explanation of
the data.    The simulation results are shown in Table~4---all of the actual
best-fit likelihoods are reasonable according to the simulations except for the
likelihood of the minimal model.   Even Model~1 passes this
goodness-of-fit test despite the strong rejection of the model by the Bayesian
model comparisons.

\begin{deluxetable}{rcccc}
\tablecolumns{5}
\tablenum{4}
\tablecaption{Goodness-of-Fit Tests via Simulations}
\tablehead{
  &           & Mean of     &  Standard   &  Fraction of  \nl
  &  Best Fit & Simulations &  Deviation  &  Simulations with  \nl
\colhead{Model} & $\log_{10} L_{\rm fit}$ & $\log_{10} L_{\rm simu}$ &
of $\log_{10} L_{\rm simu}$ &   $L_{\rm simu} < L_{\rm fit}$  }
\startdata
  0  &  161.5  &  293.8  &  6.5  &  0/1000  \nl
  1  &  211.5  &  214.9  &  6.3  &   0.29  \nl
  2  &  220.8  &  221.6  &  9.0  &   0.46  \nl
  4  &  212.6  &  213.4  &  6.2  &   0.45  \nl
 10  &  223.7  &  222.2  &  8.5  &   0.58  \nl
 12  &  222.9  &  221.0  &  8.2  &   0.60  \nl
 33  &  223.6  &  222.8  &  8.8  &   0.54  \nl
\enddata
\end{deluxetable}

\paragraph{{\bf Recommendations}.}
Models~2 and 33 both make excellent fits to the data (Tables 2 and 4)
and the two model histograms are extremely similar (Fig.~5 and 7).
While there is evidence that Model~33 is better, for most applications
we recommend Model~2 because of its simplicity.
For critical analysis, Model~33 might be selected.
The models are specified by eqs.~4 to~10.
The best-fit parameter values are listed in Table 3.
If only confidence radii are needed, these may be read from Fig.~8.
Because the models are empirically determined using 411 reference locations,
confidence radii much beyond 99\% have the character of an extrapolation and
should be used with caution.

\begin{figure*}[tbp!]
\begin{center}
\mbox{
\hspace{12mm}
\psfig{figure=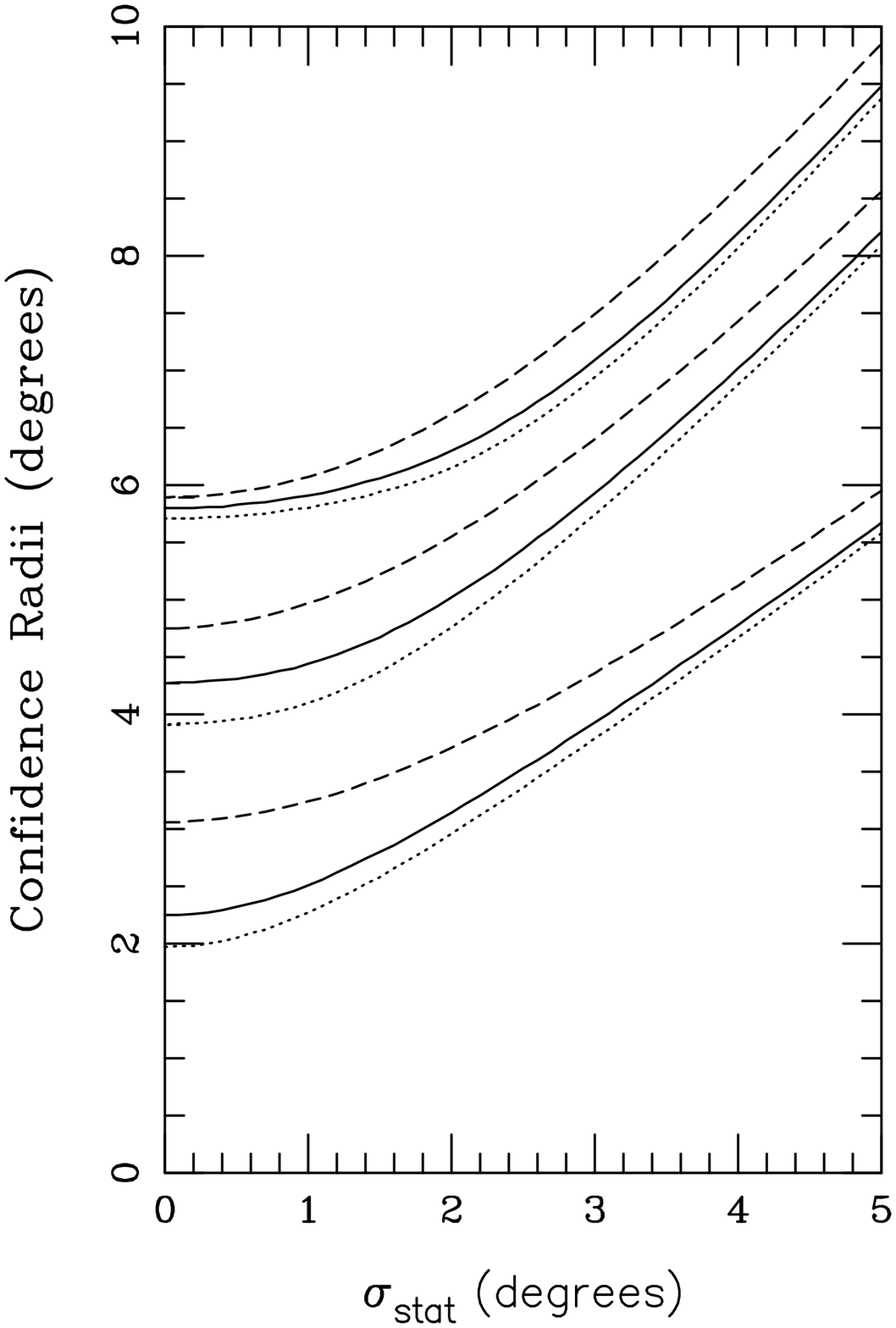,width=55mm}
\hspace{16mm}
\psfig{figure=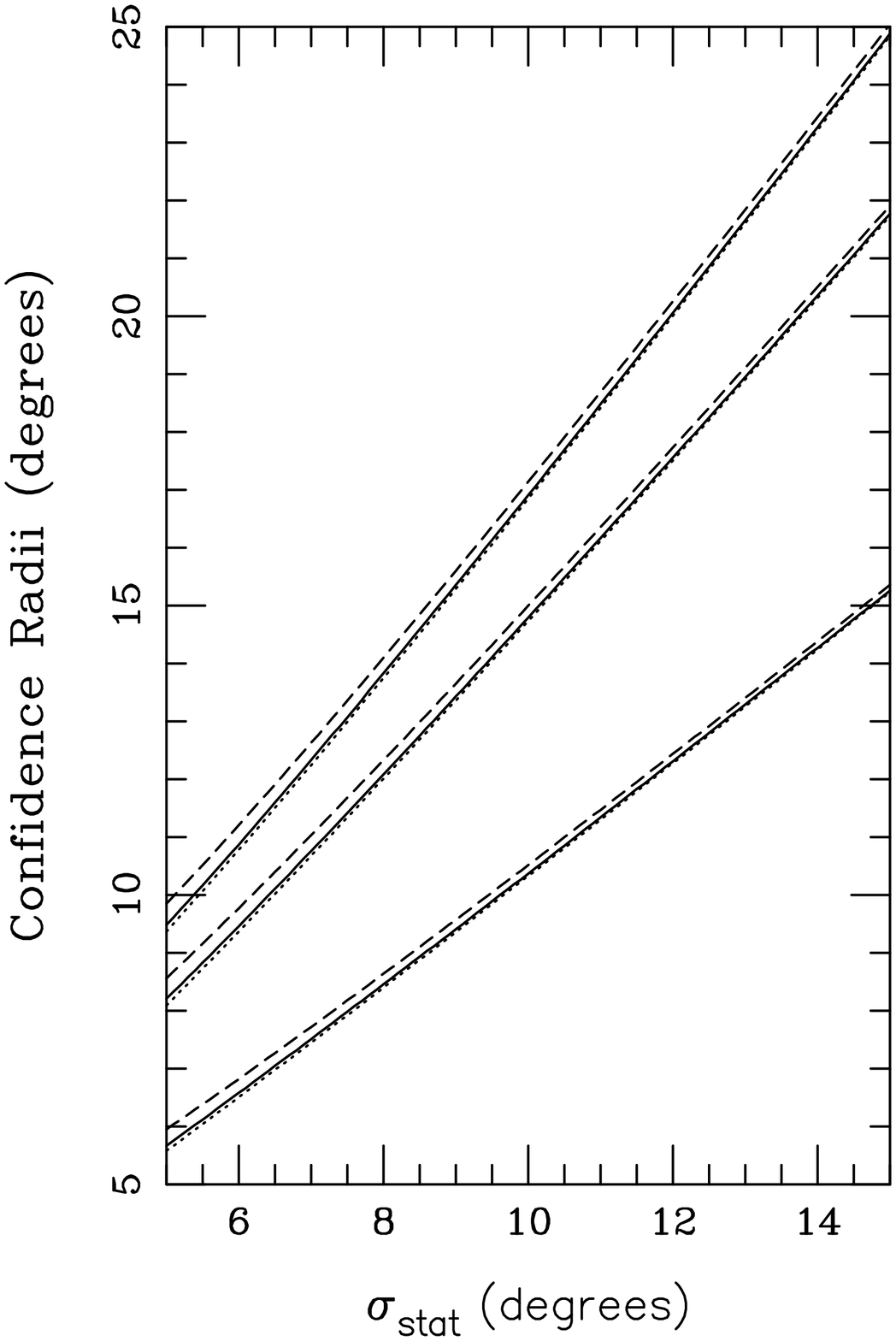,width=55mm}
\hspace{12mm}
}
\end{center}
\caption{Confidence Radii for GRB location as a function of $\sigma_{\rm stat}$
listed in the 4Br catalog.
The curves are based upon Models~2 and 33, as specified in
Table~3.
The solid curves show the confidence radii for Model~2, the dotted curves
the confidence radii according to Model~33 for locations obtained using
the CONT datatype, and the dashed curves the radii according to Model~33
for locations obtained using datatypes other than CONT.
The bottom trio of curves shows the radii in which the location will fall
68\% of the time according to the models, the middle trio the 90\%
confidence radii, and the top trio the 95\% confidence radii.}
\end{figure*}

\subsection{Comparisons with Previous Analyses}

In a preliminary version of this project 
\markcite{Bri98} (Briggs et al. 1998), similar
results were obtained.
The largest difference is that Model~10 was not presented in that work
because it was inferior to Model~12 and Model~33 was not presented because
it had not yet been tried.   Model~10 has two values of
$\sigma_{\rm sys}$ depending on datatype (6 parameters total), while
Model~12 has two values of $\sigma_{\rm sys}$ for CONT-based locations,
but only one value for non-CONT locations (4 parameters total).
Model~33 has two values for the core systematic error, but a single value
for the fraction and a single-value for the tail systematic error.
The current analysis (Table~2) shows Model~33 to be better than Model~12 by a
small odds ratio of 8.
At this level additional data or different priors might cause the model
identified as best to change--this also applies to closely related models not
presented.
This is not a problem because these three models are based upon the same
fundamental ideas (core-plus-tail distribution
and datatype dependence) and make nearly
identical predictions for location errors.

Reasons to prefer Model~10  are
that it seems very plausible that the error distribution
of both CONT and non-CONT locations should be best-fit with a
core-plus-tail model and there is
no reason that the fractions and tail systematic errors
must be identical.
(This assertion of reasonableness might be termed a ``prior'', but
we chose not to quantify our inclination.)
However, we have recommended Model~33 because it has the best odds ratio and
is simpler than Models~10 and 12; in any case these three models provide very
similar predictions.

The parameter values of the models are also slightly changed from 
the preliminary analysis.
The differences between the analyses are due to improvements
in the IPN locations between the preliminary data releases and the
published values, improvements 
in the BATSE locations between the 4B and 4Br catalog 
\markcite{Pac98} (Paciesas et al. 1998) (also see \S3.3) and
better rejection of long IPN error boxes.
The removal of a small number of outliers due to mistakes had a 
disproportionate effect on the error analysis.

An earlier work \markcite{GL96}
(Graziani and Lamb 1996), which introduced the analysis
method used herein, reached substantially different conclusions.
Their best models were intensity dependent, with $\sigma_{\rm sys}$
a power-law function of $\sigma_{\rm stat}$ (Model~4) or of
the fluence of the burst (Model~8).
Table~2 shows Model~2 to be superior to Model~4 by an odds ratio of
$3 \times 10^7$ and the histograms comparing the observations and
Model~4 (Fig.~9) show the fit to be poor.
The fluence-based model cannot be analyzed for the full 4Br catalog because
fluence values are unavailable for some bursts due to telemetry gaps that
are uncorrelated with burst properties.
Of the 411 bursts of our comparison sample, 324 have fluence measurements.
Table~5 lists model comparisons using these 324 GRBs.
Model~2 is superior to
Model~8 by an odds ratio of $1 \times 10^8$.
The parameter values found for Model~8 are
$A = 2.66^\circ \pm 0.12^\circ$ and $\alpha = -0.084 \pm 0.036$.
Like \markcite{GL96}
Graziani and Lamb (1996), we find the intensity-dependent models
to be favored over Model~1 (Tables~2 and~5).

\begin{figure*}[tbp!]
\begin{center}
\mbox{
\hspace{12mm}
\psfig{figure=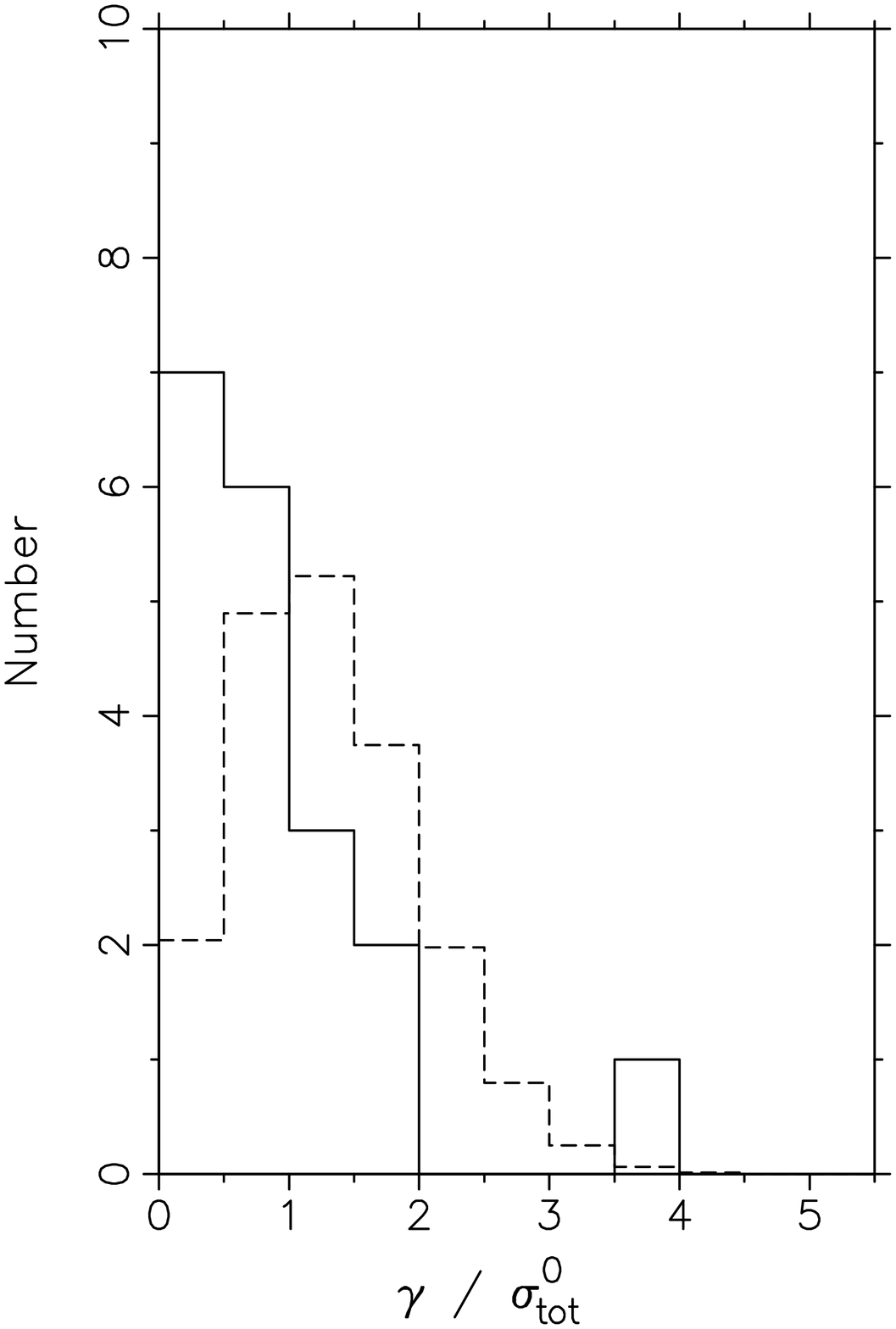,width=55mm}
\hspace{16mm}
\psfig{figure=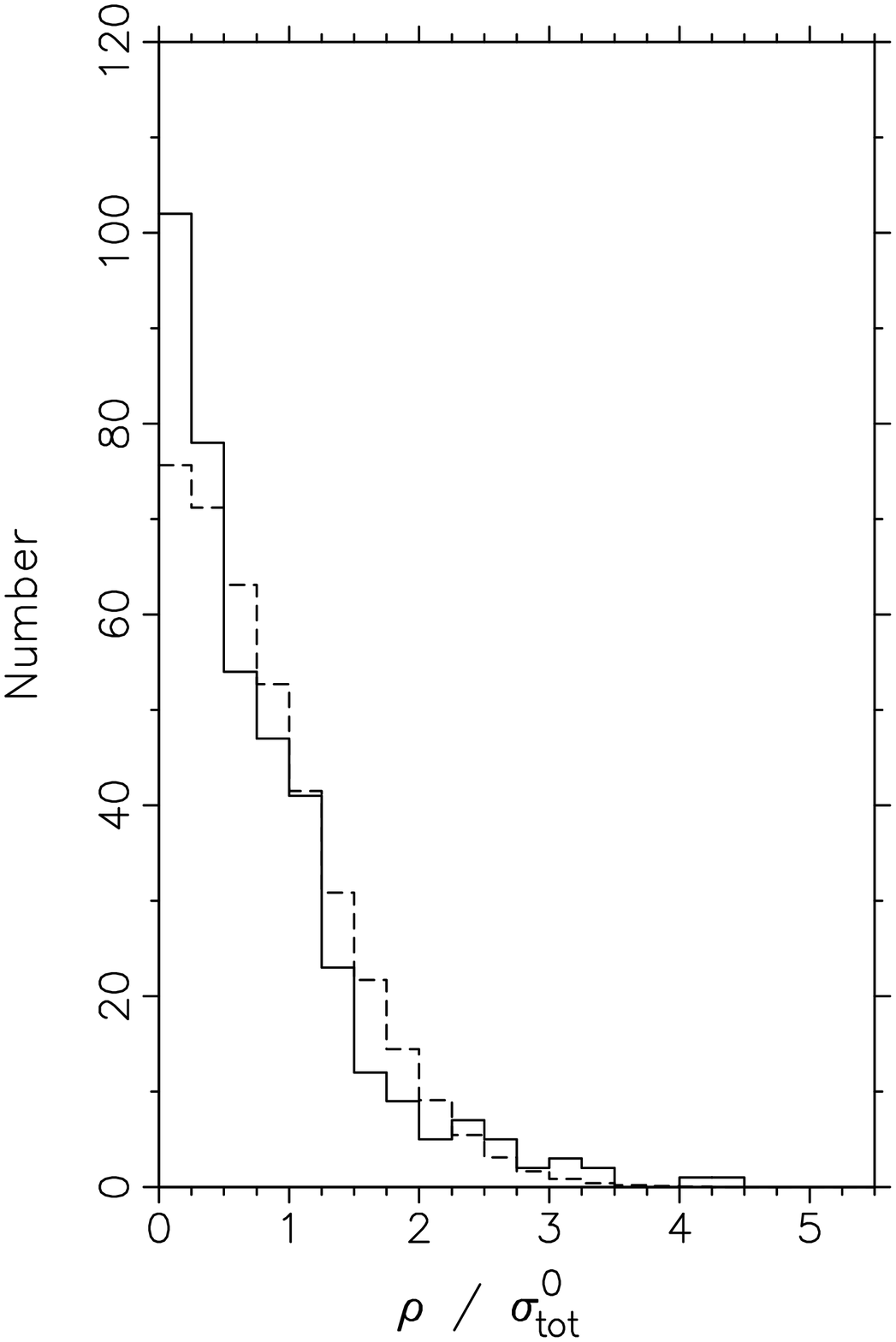,width=55mm}
\hspace{12mm}
}
\end{center}
\caption{Histograms of the data (solid) and Model~4 (dashed).
Model~4  predicts too few events at both small
($\rho / \sigma_{\rm tot}^0 \lesssim 0.25$) and large
($\rho / \sigma_{\rm tot}^0 \gtrsim 3$) separations.
}
\end{figure*}

\begin{deluxetable}{ccccc}
\tablecolumns{5}
\tablenum{5}        
\tablecaption{Models Comparison Results for 4Br Events with Fluence Data}
\tablehead{
  &   &    Number of  &  $\log_{10}$   &   $\log_{10}$  \nl
  Model  &   Description    &  
     Parameters  &  likelihood &  Odds Factor }
\startdata
   0    &
\begin{minipage}[t]{45mm}
\begin{flushleft}
Minimal: Systematic Error
Assumed to be $1.6^\circ$
\end{flushleft}
\end{minipage}  &
 0  &  134.0  &  134.0  \nl
\tablevspace{3mm}
   1    & 
\begin{minipage}[t]{45mm}
\begin{flushleft}
Single Systematic Error applied to all locations
\end{flushleft}
\end{minipage} &
1 &  168.2  &  166.9  \nl
\tablevspace{3mm}
    2   &
\begin{minipage}[t]{45mm}
\begin{flushleft}
Core-plus-Tail Systematic Errors
\end{flushleft}
\end{minipage} &
3  &    177.8  &  175.0   \nl
\tablevspace{3mm}
   4   &
\begin{minipage}[t]{45mm}
\begin{flushleft}
Systematic Error a power law
function of statistical uncertainty:
$\sigma_{\rm sys} = A (\sigma_{\rm stat} / 1^\circ)^\alpha$
\end{flushleft}
\end{minipage}    &
2  &  170.4  &  168.1  \nl
\tablevspace{3mm}
   8   &
\begin{minipage}[t]{45mm}
\begin{flushleft}
Systematic Error a power law
function of fluence:
$\sigma_{\rm sys} = A (\sigma_{\rm stat} / 1 \times 10^{-5})^\alpha$
\end{flushleft}
\end{minipage}    &
2  &  169.4  &  166.8  \nl
\enddata
\end{deluxetable}

Graziani and Lamb \markcite{GL96} (1996) analyzed the 
3B catalog \markcite{Mee96}
(Meegan et al. 1996) (the 4Br catalog did not yet exist) and
restricted the analysis to events with single annuli.
They neglected all of the events with $\gamma$ measurements
in order to avoid any locations that might have been used to optimize
the location program LOCBURST.
Twelve GRBs with $\gamma$-values were used to identify
which proposed algorithmic improvements were worth
implementing \markcite{Pen98} (Pendleton et al. 1998).   
All location algorithm parameter values were determined
from laboratory measurements, in-orbit observations of
the Crab Nebula and pulsar and from Monte Carlo simulations--none
were optimized using GRB measurements.
We know of no reason to exclude all $\gamma$ measurements and judge that
there is little `circularity' in using the $\gamma$ measurements of
the twelve `study' bursts.
In any case, excluding the data for these twelve events has
negligible effects on the results.

The reasons that are probably most important in explaining the difference
between our results and those of Graziani and Lamb \markcite {GL96} (1996) are 
the models tested and revisions to the preliminary IPN  data
\markcite{KH98a} (Hurley et al. 1998a).
Using the original 3B BATSE data and the data selections of Graziani and
Lamb (i.e., excluding bursts with $\gamma$ measurements), but using
the finalized IPN data,
Model~2 is favored over Model~4 by an odds ratio of 400.
Adding the $\gamma$ measurements for all except the twelve `study' bursts,
the odds ratio becomes 4700.
Unlike Graziani and Lamb, we conclude that there is no evidence for a direct
dependence of BATSE systematic location errors on burst intensity.

Because the IPN Supplement emphasizes relatively bright BATSE bursts
(Fig.~2), it is difficult to use IPN data to test the locations of faint
bursts.  
Graziani and Lamb \markcite{GL96} (1996) used extrapolations of Models~4 and~8 
to argue that faint BATSE
bursts have large systematic errors, e.g., $\sigma_{\rm sys} = 8^\circ$ for
bursts with $\sigma_{\rm stat}= 10^\circ$,
and that the 3B catalog had larger
systematic errors for faint bursts than the 1B catalog.
Because our analysis finds no evidence that $\sigma_{\rm sys}$ directly depends
on burst intensity, we consider such extrapolations unjustified.

A more direct test of the location accuracy of weak bursts can be made using
fluctuations from Cygnus X-1 which triggered BATSE.
These fluctuations are almost always near the trigger threshold.
The spectra of these fluctuations are similar to GRBs in the energy
range (50--300 keV) used for determining locations, but softer at
higher energies.
Because of the reduced flux above the energy range used for burst locations,
the scattering model is less important for locating Cygnus X-1
than for GRBs; however, any
intensity dependence of scattering other than the obvious proportionality
is  implausible.
Thirty-nine such events had an average statistical uncertainty
$\sigma_{\rm stat}$
of $12^\circ$, but an average separation from the true location
of only $10^\circ$, implying a 95\% confidence upper-limit on
$\sigma_{\rm sys}$ of $7^\circ$ \markcite{Mee96} (Meegan et al. 1996).

There is, however, the possibility of a weak {\it indirect} correlation
of $\sigma_{\rm sys}$ with burst intensity.    
The fraction of events located using CONT data decreases with decreasing
burst intensity (Fig.~10).
The reasons for this trend are correlations between the
optimum BATSE datatype for a location, GRB duration, and $\sigma_{\rm stat}$:
short events are best located with datatypes other than CONT to avoid adding
background intervals to the burst data; additionally 
short events typically have low fluences and thus larger values
of $\sigma_{\rm stat}$.
In Model~33 the only difference between locations obtained using CONT data and
other locations is the value of the core systematic error.
Coupled with the correlation between datatype and intensity, this datatype
dependence of Model~33 implies that a larger fraction of
weak bursts have a core systematic error of
$2.74^\circ$ instead of $1.67^\circ$.
This difference is unimportant for bursts with large
values of $\sigma_{\rm stat}$.
The systematic error is not a direct function of the statistical
uncertainty and therefore does not extrapolate to large values for
faint bursts.

\begin{figure*}[tbp!]
\begin{center}
\mbox{
\psfig{figure=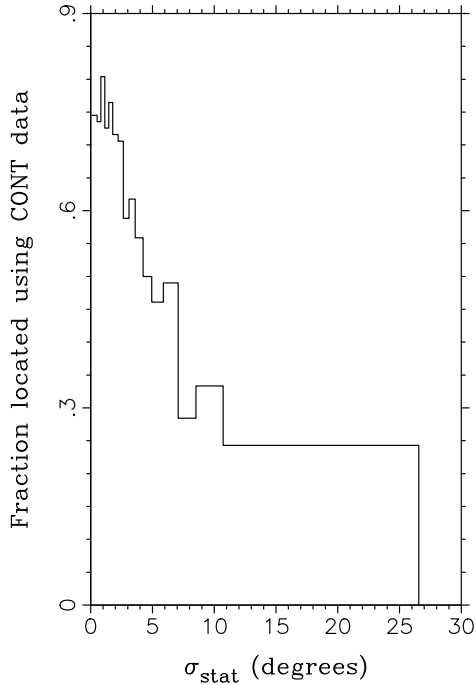,width=65mm}
\hspace{3mm}
\begin{minipage}[b]{90mm}
\caption{Fraction of events located using the CONT datatype versus
$\sigma_{\rm stat}$.   Each of the first 15 bins has 102 locations, while
the last bin has 107.}
\protect \vspace{45mm}
\end{minipage}
}
\end{center}
\end{figure*}

It seems plausible that the improvement of the models with
direct intensity dependencies (Models~4 and~8) over
single-parameter models (0~or~1) is due to two correlations: the
correlation between
intensity and datatype used for the location
and the probable correlation between
datatype and systematic error.
However, this may not be be the case since a model in which
$\sigma_{\rm sys}$ has no dependence on any burst property  (Model~2)
is favored over Model~4 by a large odds ratio.

\subsection{Previous BATSE Catalogs}

The locations of the previous BATSE catalogs, 
1B \markcite{Fish94} (Fishman et al. 1994), 
2B \markcite{Mee94} (Meegan et al. 1994),
3B \markcite{Mee95} (Meegan et al. 1995), 
and 4B \markcite{Pac97} (Paciesas et al. 1997) are superseded
by the revised locations of the 4Br catalog 
\markcite{Pac98} (Paciesas et al. 1998), with the exception that
bursts of the 3B catalog which are not listed in the 4Br catalog paper
have unaltered locations which are considered to be 4Br locations.
Key events affecting the quality of the locations
were the deterioration of the data availability starting at the
end of the 1B catalog due to the failure of the tape recorders on-board CGRO,
a substantial recovery of data availability at the end of the 2B catalog
due to flight software improvements, and a major improvement in
the location algorithm effective with the publication of the 3B catalog
(at which time all locations were revised).
A lesser algorithmic improvement is implemented with the publication of
the 4Br catalog
\markcite{Pac98} (Paciesas et al. 1998), in which 208 locations are
changed from the original 4B catalog.

Table~6 presents analyses of these catalogs for
historical interest and to aid in assessing work based upon these catalogs.
Since the purpose is to understand BATSE location errors, 
in all cases the current IPN Supplement data are used rather than the
preliminary IPN data available at the time of the publication of
a particular BATSE catalog.
Only Models~1 and~2  are shown because
most of the catalogs are too small to support reliable analysis
of models with more than about 3 parameters.
The odds ratios favor Model~2 over Model~1 by factors
ranging from 4 to $3 \times 10^{13}$.
The major algorithmic upgrade of LOCBURST can be seen in the comparison
of the parameter values for the original 2B and 3B catalogs, or between
the original 2B catalog and the equivalent subset of the 4Br catalog.
The minor algorithmic improvement between the 4B and 4Br catalogs is
not apparent in Table~6, which is not surprising since only 13\% of the 
locations were altered, mostly modestly.
A more direct comparison validates the  algorithmic upgrade used for
the 4Br catalog \markcite{Pac98} (Paciesas et al. 1998).
There is no evidence in the parameter values for subsets of the 4Br
catalog for any time dependence of location quality, especially considering
that the parameter values of Model~2 are correlated.
In particular, the locations obtained during the period of lesser-quality
data (2B$-$1B) have error model parameters consistent with the other
time periods.

\begin{deluxetable}{crrrrrcccc}
\footnotesize
\tablecolumns{10}
\tablenum{6}
\tablecaption{Error Models for Various Datasets}
\tablehead{
    & &  &  \colhead{Num. of}   &  
 \colhead{Num.}  &  \colhead{Num.}  & 
    Model 1  & \multicolumn{3}{c}{Model 2}  \nl
    & \multicolumn{2}{c}{Trigger} &  \colhead{BATSE} & 
    \colhead{w. IPN}    & \colhead{with}     & 
   $\sigma_{\rm sys}$ & 
         $\sigma^1_{\rm sys}$ &
   $f_1$  &  $\sigma^2_{\rm sys}$ \nl
Catalog & \multicolumn{2}{c}{Range}  & \colhead{GRBs}  &  \colhead{data} &
  \colhead{$\gamma$}  &  (deg)     &   (deg) &  &  (deg)  \nl
  }
\startdata
\cutinhead{Superseded catalogs analyzed using originally published data:}
 1B   &   105 & 1466 &  260 &  54 &  6 &    3.4 $\pm$ 0.4  &  
   1.6 $\pm$ 0.4 & 0.57 $^{+0.15}_{-0.19}$&  $5.0^{+1.3}_{-0.9}$ \nl
 2B   &   105 & 2230 &  585 & 128 & 18 &    4.5 $\pm$ 0.3  &  
   1.8 $\pm$ 0.3 & 0.75 $\pm$ 0.07 &  $9.3^{+1.7}_{-1.4}$ \nl
 3B   &   105 & 3174 & 1122 & 264 & 19 &  2.88 $\pm$ 0.15 & 
  1.69 $\pm$ 0.22 & 0.71 $\pm$ 0.11 & 4.9 $\pm$ 0.8 \nl
 4B   &   105 & 5586 & 1637 & 411 & 19 &  2.90 $\pm$ 0.12 &
  1.84 $\pm$ 0.18 & 0.74 $\pm$ 0.09 &  5.0 $\pm$ 0.7 \nl
\cutinhead{Subsets of the 4Br catalog:}
 full   &   105  & 5586 & 1637 & 411 & 19 &  2.80 $\pm$ 0.12 &
  1.85 $\pm$ 0.16 & $0.78 \pm 0.08$ &  $5.1^{+0.8}_{-0.6}$ \nl
 1B   &   105  & 1466 &  263 &  54 &  6 &  2.56 $\pm$ 0.30 & 
  $1.6 \pm 0.3$ &  $0.79^{+0.10}_{-0.20}$  &  $5.4^{+2.5}_{-1.5}$ \nl
 2B   &   105  & 2230 &  586 & 128 & 18 &  2.80 $\pm$ 0.20 &  
  1.55 $\pm$ 0.26 &  $0.72^{+0.09}_{-0.13}$ &  $5.2^{+1.2}_{-0.9}$ \nl
2B$-$1B & 1467 & 2230 &  323 &  74 & 12 &  3.0 $\pm$ 0.3   &  
  1.5 $\pm$ 0.4  & 0.67 $\pm$ 0.18 &   $5.1^{+1.5}_{-1.0}$ \nl
 3B    &  105  & 3174 & 1122 & 264 & 19 &  2.89 $\pm$ 0.15 & 
  1.77 $\pm$ 0.18 & 0.76 $\pm$ 0.09 &  $5.4^{+0.9}_{-0.7}$ \nl
4B$-$2B & 2232 & 5586 & 1051 & 283 &  1 &  2.80 $\pm$ 0.14 & 
   2.08 $\pm$ 0.21 & $0.84^{+0.07}_{-0.12}$ &  $5.4^{+1.3}_{-1.0}$ \nl
4B$-$3B & 3177 & 5586 &  516 & 147 &  0 &  2.61 $\pm$ 0.20 &  
  2.1 $\pm$ 0.4  & $0.81^{+0.16}_{-0.35}$ &  $4.4^{+3.8}_{-1.2}$ \nl
\enddata
\end{deluxetable}

\section{Conclusions}

This paper presents an improved model of the BATSE GRB location errors.
Future work may incorporate improvements such as
the incorporation of more locations (BATSE and IPN), 
additional `point' IPN locations when more interplanetary spacecraft,
such as the Near Earth Asteroid Rendezvous (NEAR) mission, 
are added to the network, the use of locations obtained with BeppoSAX (which was
launched several months before the end of the 
4Br catalog) and HETE~II, and
the extension of the analysis to use comparison locations with errors
not negligible compared to BATSE (e.g., COMPTEL and WATCH locations).
Another possible extension would be to use elliptical error boxes rather
than the azimuthally symmetric distributions assumed here.
We continue to search for correlations between location errors and
other properties.   Such correlations might explain why some bursts are
in the tail of the location distribution.
If we identify a correlation, we might be able to use that knowledge
to improve the location algorithm.

The parameter values of model fits to catalogs preceding the 4Br catalog
confirm that the location algorithm introduced with the 3B catalog
was a significant improvement over the previous  algorithm.

The models of this paper are highly significant improvements
over previous models of the BATSE GRB location error distribution.
We recommend the three-parameter `core-plus-tail' model (Model~2) for most
analyses.   There is evidence that a `core-plus-tail' model with
core systematic error depending on datatype (Model~33) is superior, and this
model can be used for the most critical applications.

A promising use of BATSE and IPN locations is rapid (minutes to hours)
optical observations of GRB locations obtained with the BATSE
Rapid Burst Response (RBR) System 
\markcite{Kip98} (Kippen et al. 1998), followed by detailed
analysis of the smaller error box obtained either with the intersection
of a single IPN annulus with the BATSE box or the very small error box
obtained by the intersection of a pair of IPN annuli.
The RBR locations are obtained by human operators using the standard
LOCBURST algorithm and have  accuracies consistent with 
the accuracies of BATSE catalog locations.

\acknowledgements

MSB, GNP, RMK, and JJB acknowledge NASA Grant NCC8-65 for support.
KH is grateful to JPL Contract 958056 for support of Ulysses operations
and to NASA Grant NAG5-1560 for support of the IPN.    We appreciate the
suggestions of the anonymous referee.

\clearpage
         
\appendix
\section{Derivation of $p(\rho)$}

The goal is to test the probability density function $p(\gamma)$ for the
separation $\gamma$ between the BATSE location and the true location.
In most cases only single IPN annuli are available and
therefore only the smallest distances $\rho$ between the BATSE location and the
IPN annuli can be determined.
In such cases
the testing of $p(\gamma)$ must be done indirectly by comparing
$p(\rho)$ to the observations.
The subject of this Appendix is deriving $p(\rho)$ given
$p(\gamma)$, assumed to be the
Fisher distribution (eq.~5).

Figure 11 depicts the geometry of comparing a BATSE location to a single
IPN annulus.  
All circles and positions are on the surface of the unit sphere.
The IPN annulus, which may not be a great circle, has radius
$r$ and is assumed to have a negligible width $\delta r$.  
The coordinate system is chosen so that the center of the IPN
annulus is at $\hat  {z}$ and the true location $\hat  {T}$
is in the $xz$ plane.   The BATSE location $\hat {B}$
is at an arbitrary position on the unit sphere.

\begin{figure*}[bp!]
\begin{center}
\mbox{
\psfig{figure=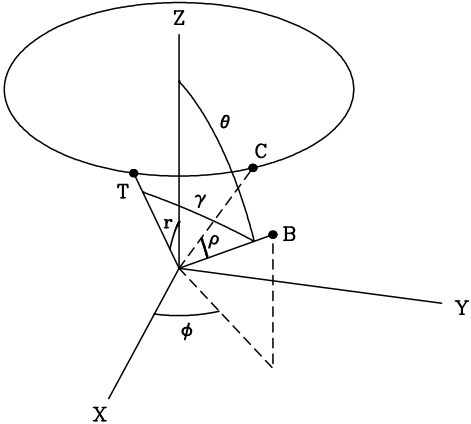,width=65mm}
\hspace{3mm}
\begin{minipage}[b]{90mm}
\caption{Diagram of the geometry relating the true location $\hat  {T}$
to the BATSE location $\hat  {B}$.   The coordinate system is
chosen so that the center of the IPN annulus is at $\hat  {z}$
and the true location $\hat  {T}$ is in the xz-plane.
The annulus has radius $r$.
The BATSE location $\hat  {B}$ is at spherical coordinates
$\theta, \phi$.    The angle between the true location $\hat  {T}$
and the BATSE location $\hat  {B}$ is $\gamma$, 
while the angle between the BATSE location 
$\hat  {B}$ and the closest location on the
annulus $\hat  {C}$ is $\rho$.}
\protect \vspace{10mm}
\end{minipage}
}
\end{center}
\end{figure*}

Locations may be expressed in spherical coordinates
$(\theta, \phi)$, where $\theta$ is the  angle from $\hat {z}$
and $\phi$ is the azimuthal angle from $\hat {x}$ (see Fig.~11).
Since
\begin{equation}
\rho = | \theta - r |,
\end{equation}
the probability density function $p(\rho)$ can be readily obtained from
$p(\theta)$.
The assumed probability density function is a function of $\gamma$, so it is
necessary to derive the the relationship between the
the coordinate system $(\theta, \phi)$ and the system
$(\gamma, \psi)$, where $\gamma$ is the  angle
from $\hat  {T}$ to $\hat  {B}$
and $\psi$ is an azimuthal angle about $\hat  {T}$.

\begin{equation}
\hat  {T} = \sin{r} \: \hat  {x} +
\cos{r} \: \hat  {z},
\end{equation}

\begin{equation}
\hat  {B} =
\sin{\theta} \cos {\phi} \: \hat  {x} +
\sin{\theta} \sin {\phi} \: \hat  {y} +
\cos{\theta} \:  \hat  {z}.
\end{equation}

Therefore,
\begin{equation}
\cos{\gamma} = \hat  {T} \cdot \hat  {B}
\end{equation}
\begin{equation}
= \cos (\theta - r) - \sin {\theta} \sin{r} (1 - \cos {\phi}).
\end{equation}

Since the transformation is a rotation, the Jacobean determinant is
unity and the transformation between probability density functions is simply
$p(\theta, \phi) = p(\gamma, \psi)$.
The probability density function is assumed to be
rotationally symmetric,
so $p(\gamma, \psi) \equiv p(\gamma)$.

Substituting for $\cos{\gamma}$ from eq.~A5 into eq.~5 for
the Fisher distribution, 
\begin{equation}
p(\theta, \phi) \; d \Omega
= \frac{\kappa}{2 \pi ({\rm e}^{\kappa} -{\rm e}^{-\kappa})} 
{\rm e}^{\kappa \cos{(\theta - r)}}
{\rm e}^{-\kappa \sin{\theta} \sin{r} (1-\cos{\phi})} \; d \Omega
\end{equation}
and
\begin{equation}
p(\theta) \; d \theta = \int_{0}^{2 \pi} d \phi \: p(\theta, \phi) \;\: d \theta
\end{equation}
\begin{equation}
= \frac{\kappa}{\pi ({\rm e}^\kappa - {\rm e}^{-\kappa})}
{\rm e}^{\kappa \cos{(\theta - r)}}
\int_{0}^{\pi} d \phi \:
{\rm e}^{-\kappa \sin{\theta} \sin{r} (1 - \cos {\phi})} \;\: d \theta .
\end{equation}
Finally, using formula 8.431.3 of Gradshteyn and Ryzhik \markcite{GR80} (1980),
\begin{equation}
p(\theta) \; d \theta
= \frac{\kappa}{{\rm e}^\kappa - {\rm e}^{-\kappa}}
{\rm e}^{\kappa \cos{(\theta - r)}}
\frac{I_0 (\kappa \sin{\theta}\sin{r})} {{\rm e}^{\kappa \sin{\theta} \sin{r}}}
\; d \theta .
\end{equation}

Including the factor ${\rm e}^{\kappa \sin{\theta} \sin{r}}$ in the last term 
makes that term  more amenable to numerical evaluation (e.g.,
routine BESSI0 of Press et al. \markcite{NR} (1992) 
is easily modified in this manner).

Usually there are two values of $\theta$ for a value of $\rho$, 
corresponding to locations inside and outside of the annulus, so that
\begin{equation}
p(\rho) \; d \rho
= p(\theta = r - \rho) \; d \theta +
  p(\theta = r + \rho) \; d \theta.
\end{equation}
In the unusual case that $\rho > r$, only the $\theta = r + \rho$
term contributes.


\end{document}